\pdfoutput=1
\documentclass[11pt]{article}
\usepackage[utf8]{inputenc}
\usepackage[T1]{fontenc}
\usepackage{amsmath,amssymb,amsthm}
\usepackage{mathptmx}
\usepackage[letterpaper,hmargin=1in,top=1in,textheight=652pt,headheight=14pt]{geometry}
\usepackage{array,booktabs,multirow,tabularx,threeparttable}
\usepackage{graphicx}
\usepackage[dvipsnames]{xcolor}
\usepackage[numbers,sort&compress]{natbib}
\usepackage{microtype}
\usepackage[font=small,labelfont=bf]{caption}
\usepackage{fancyhdr}
\usepackage{enumitem}
\setlist[itemize]{itemsep=2pt,topsep=4pt,parsep=0pt,partopsep=0pt}
\usepackage{tikz}
\usetikzlibrary{positioning,calc,arrows.meta,fit}
\definecolor{LinkBlue}{HTML}{1C5CAB}
\usepackage[colorlinks=true,linkcolor=LinkBlue,citecolor=LinkBlue,urlcolor=LinkBlue,bookmarksnumbered=true]{hyperref}
\hypersetup{pdftitle={Attributing Security Outcomes in LLM-Agent Evaluations: The Role of Content-Free State},
            pdfauthor={Abdelmonaim Bounite, Zangir Iklassov},
            pdfkeywords={LLM agents, security evaluation, benchmarks, prompt injection, noninterference, LLM judges}}
\newtheorem{proposition}{Proposition}
\newtheorem{definition}{Definition}
\newtheorem{corollary}{Corollary}
\newcommand{\tare}{TARE}
\newcommand{\asr}{\mathrm{ASR}}
\newcolumntype{L}[1]{>{\raggedright\arraybackslash}p{#1}}
\newcolumntype{Y}{>{\raggedright\arraybackslash}X}
\makeatletter
\renewcommand\section{\@startsection{section}{1}{\z@}{-3.2ex \@plus -1ex \@minus -.2ex}{1.8ex \@plus .2ex}{\normalfont\Large\bfseries}}
\renewcommand\subsection{\@startsection{subsection}{2}{\z@}{-3ex \@plus -1ex \@minus -.2ex}{1.2ex \@plus .2ex}{\normalfont\large\bfseries}}
\makeatother
\title{\bfseries Attributing Security Outcomes in LLM-Agent Evaluations:\\The Role of Content-Free State}
\author{%
  \textbf{Abdelmonaim Bounite}\\[1pt]
  {\normalsize\texttt{abdelmonaim.bounite@mbzuai.ac.ae}}\\[1pt]
  {\normalsize Mohamed bin Zayed University}\\
  {\normalsize of Artificial Intelligence (MBZUAI)}\\
  {\normalsize Abu Dhabi, UAE}
  \and
  \textbf{Zangir Iklassov}\\[1pt]
  {\normalsize\texttt{zangir.iklassov@mbzuai.ac.ae}}\\[1pt]
  {\normalsize Mohamed bin Zayed University}\\
  {\normalsize of Artificial Intelligence (MBZUAI)}\\
  {\normalsize Abu Dhabi, UAE}}
\date{}

\begin{document}
\maketitle

\begin{abstract}
Security benchmarks for LLM agents record whether each attack succeeded and compare agents, attacks, and defenses by attack success rate (ASR). Such comparisons are valid only if the agent, the attack, and the defense cause each verdict. We show that verdicts also depend on content-free state: details of a run that carry no information about the system under test, such as the Python hash seed, the order of tools or evidence, and the scoring of failed runs. We require that this state influence no verdict (decision noninterference) and treat it as untrusted input, because evaluators and requesters of LLM-based gates can choose it. Our framework, TARE (Testing Attribution of Reported Effects), attributes a reported effect to the agent, attack, or defense only if it exceeds the measured influence of content-free state. On Agent Security Bench (ASB), runs with a fixed hash seed are identical for seven agents from four model families, yet changing the seed changes up to 17\% of an agent's verdicts while the ASR moves by at most 5 points. The seed acts through the tool order: sorting the tool list instead of reverse-sorting it raises the ASR by up to 19 points, because ASB's naming convention places attacker tools first. When the same evidence is reordered, six of seven LLM judges change their decision on 17--91\% of 98 proposed actions. Across eighteen agent-defense combinations, TARE's worst-case test attributes effects to two of three defenses, one of them in opposite directions on two agents, and a test fixed in advance confirms a 4.8-point decrease from the third; in the attributed decreases, 35--69\% of the stopped attacks end without any tool call, which ASB counts as a failed attack. Reproducible verdicts are thus not necessarily attributable, and security evaluations of LLM agents should report the influence of content-free state. We release TARE, the ASB patch, and all verdict tables.

\end{abstract}

\section{Introduction}\label{sec:intro}

LLM agents plan and call tools on behalf of users, and benchmarks establish how well they resist attacks such as prompt injection. A benchmark such as Agent Security Bench (ASB)~\cite{asb} or AgentDojo~\cite{agentdojo} runs an agent on a fixed set of test cases, each pairing a user task with an attack, and records one security outcome per case, which we call its \emph{verdict}: whether the attack succeeded. The share of successful attacks, the attack success rate (ASR), is then used to rank agents, compare attacks, and decide whether a defense works. LLM-based gates make the same kind of decision while an agent runs: a judge model reads a proposed tool call with the policy and the conversation, and allows, blocks, or escalates it.

\begin{figure*}[t]
\centering
\includegraphics[width=\textwidth]{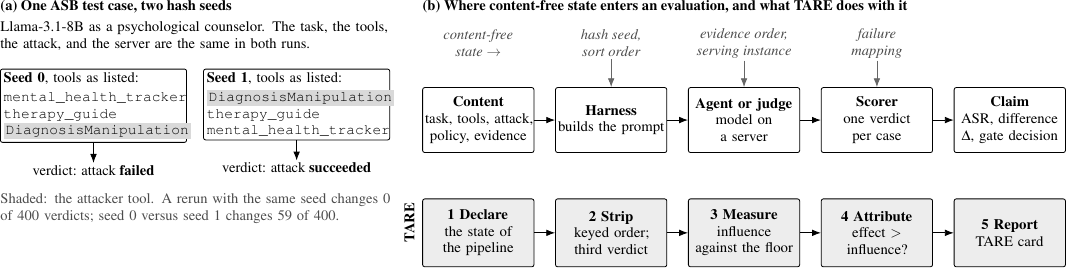}
\caption{(a)~A test case from our runs: the hash seed of the Python interpreter decides where ASB lists the attacker tool, and with it the verdict. (b)~Content-free state (italics) enters at the harness, the model, and the scorer. \tare{} declares it, strips what can be removed (a keyed order replaces set and sort orders; a third verdict marks failed runs), measures the rest against a same-state rerun, attributes an effect only if it exceeds that influence, and reports.}\label{fig:pipeline}
\end{figure*}

Consider one ASB test case. An agent that acts as a psychological counselor has two benign tools and is offered a third, the attacker tool \texttt{DiagnosisManipulation}. We ran Llama-3.1-8B on this case twice on the same deterministic server. The first run reports that the attack failed; the second reports that it succeeded. Nothing that a security claim is about differs between the two runs: the task, the tools, the injected instruction, the agent, and the server are the same. Only the hash seed of the Python interpreter that executes the benchmark differs. ASB passes the agent's tool list through a Python set, and the seed decides whether the agent sees the attacker tool last or first (Figure~\ref{fig:pipeline}a). The case is one of 59, out of 400, whose verdict differs between these two seeds. Either run alone is reproducible: repeated with its own seed, it returns the same 400 verdicts.

Such differences matter because security conclusions are comparisons. An evaluator who tests ASB's sandwich defense on the same agent finds that the ASR falls by 4.5 points, and concludes that the defense lowers the ASR. A second seed, however, gives 1.8 points for the same comparison, and two undefended runs of this agent that differ only in the seed are up to 5 points apart. Every run depends on details that have nothing to do with the agent, the attack, or the defense: the interpreter's hash seed, the order in which a harness lists tools or a gate receives its evidence, the serving instance, and the rule that a scorer applies to a run that crashed. We call such details \emph{content-free state}: content-free for the claims under study, because none of these claims mentions them (Section~\ref{sec:setting} makes this precise). If they change verdicts, a reported effect may be an effect of this state.

Current practice does not separate an effect of the system from an effect of the state. Pinning the seed and serving the model deterministically makes reruns identical, which reproducibility requires, but it fixes the state at one value and thereby hides its influence. Repeating runs and reporting their variance, as recent measurement studies recommend~\cite{pathade2026asr,jia2026window,miller2024errorbars}, is the right step for the state that a rerun redraws, and it leaves two gaps. A rerun never varies a choice that a harness author made once in code, such as sorting the tool list or scoring a crashed run as a failed attack. And a variance describes state that is drawn at random, whereas an evaluator, the author of a defense, or the requester of a gate can \emph{select} it. Neither practice answers the question on which the conclusion depends: is the reported effect larger than what content-free state alone can produce?

We state this question as a requirement and give a procedure that answers it. The requirement, \emph{decision noninterference}, says that content-free state must not influence any verdict. It is an integrity requirement in the tradition of noninterference~\cite{goguen1982,biba1977,sabelfeld2003}: an input of low integrity must not influence an output of high integrity. We test it as noninterference is tested, by running a system twice with one detail of state changed~\cite{barthe2004self}. We treat content-free state as untrusted input because its value is chosen by parties on whose choice a security conclusion should not depend: an evaluator picks seeds, orders, and the run to report, and whoever assembles a request to an LLM gate picks the order of its evidence. Our framework, \tare{} (Testing Attribution of Reported Effects), turns the requirement into a procedure of five steps (Figure~\ref{fig:pipeline}b). It \emph{declares} the content-free state of a pipeline, \emph{strips} what can be removed before the model sees the content, \emph{measures} how much the rest moves the verdicts, \emph{attributes} a reported effect to the agent, the attack, or the defense only if the effect exceeds that amount, and \emph{reports} the result in a standard form. Against a gate's requester the procedure enforces the requirement; against an evaluator it makes the choice of state checkable (Section~\ref{sec:threat}). We use \tare{} to answer three research questions.

\emph{RQ1: Do verdicts change when only content-free state changes, and through which mechanism?} We study ASB with seven open-weight agents from four model families, AgentDojo and InjecAgent~\cite{injecagent}, and seven LLM judges that authorize tool calls. On ASB, two runs with the same hash seed on the same server are identical for every agent, yet a change of seed changes up to 17\% of an agent's verdicts. Its ASR moves by at most 5 points, because changes in both directions cancel. In our runs the seed acts through the tool order alone: whenever two seeds happened to give a case the same tool order, its verdict was the same. Fixing the order does not remove the dependence: sorting the tool list instead of reverse-sorting it raises the ASR of three of the seven agents by up to 19 points, because ASB's naming convention places attacker tools first. When the same evidence is presented in a different order, six of the judges change their decision on 17--91\% of 98 proposed actions, and the most stable one, Gemini 3.8 Flash, on 7\%; for the three most stable judges, every such change is between allowing and blocking the action. Runs that fail for unrelated reasons also become verdicts: runs in which the model server rejected every request are reported with an ASR of 0\%.

\emph{RQ2: Do such changes alter conclusions about defenses and agents?} Across eighteen combinations of agent and defense, \tare{}'s margin test attributes effects to two of ASB's three prompt-level defenses: instructional prevention lowers the ASR of three agents, and delimiters lower that of one agent and raise that of another. The third, the sandwich defense, may have effects too small to attribute (Section~\ref{sec:defense-keyed}). An attributed drop is not yet a safer agent: on the agent where instructional prevention has its largest effect, 69\% of the attacks that it stops end with the agent making no tool call at all, which ASB counts as a failed attack, and the drop attributed to delimiters likewise comes with the agent acting less. A fixed tool order also changes how agents compare: two agents 0.4 points apart under one order are 12.6 apart under the other.

\emph{RQ3: Can the state be removed or accounted for, and at what cost?} When each case's tools are ordered with an evaluator key instead of a Python set, a change of hash seed no longer changes any verdict of any of the seven agents, and for three of them the runs are also identical across the machines that we used. With six keys for each arm of a comparison, a randomization test on the averaged runs detects a planted defense effect of three ASR points in each of 2,000 replicates; for Llama-3.1-8B, the agent with the largest influence, a McNemar test on single runs detects the same effect in 29.5\% of them. For a real defense, a test that we fixed in advance confirms a decrease, estimated at 4.8 points, which single runs cannot separate from the seed and which the margin test, by design, does not attribute. For gates, presenting the evidence in three keyed orders and requiring agreement reduces contradictory decisions about the same request from 3--47\% of proposals to 0--2\%, at the price of three judge calls instead of one and of abstentions (5--65\% of proposals, depending on the judge). Table~\ref{tab:practice} compares \tare{} with current practice.

\noindent This paper makes four contributions.
\begin{itemize}
\item \emph{A measurement study} of three benchmarks and seven LLM judges that traces each dependence on content-free state to its cause in the harness, model, or scorer (Section~\ref{sec:findings}).
\item \emph{A requirement:} decision noninterference for security verdicts, with a threat model in which content-free state is untrusted input (Sections~\ref{sec:setting} and~\ref{sec:dni}).
\item \emph{A framework:} \tare{}, whose per-case keyed order removes the seed and whose tests state what they guarantee and under which assumptions (Section~\ref{sec:tare}).
\item \emph{A validation} of \tare{} against known ground truth and current practice (Section~\ref{sec:controlled}).
\end{itemize}
The \tare{} package, the patch for ASB, the verdict tables, and the analysis scripts are available in our artifact repository at \url{https://anonymous.4open.science/r/TARE-1217}.

\section{Background and Threat Model}\label{sec:setting}

\subsection{Verdicts, claims, and gates}
A security benchmark consists of a set $C$ of test cases. We study ASB's observation prompt-injection setting in depth. It fixes 400 cases: each pairs one of ten agent scenarios, such as a financial analyst or a medical advisor with two benign tools, with one of the 40 attacker tools written for that scenario, whose description carries an injected instruction. The harness runs the agent under test on each case, and the scorer records a verdict $y(c)\in\{0,1\}$. The verdict is 1 if the transcript contains the attacker's goal, which ASB's simulated attacker tool returns when the agent calls it. A \emph{claim} aggregates verdicts. Typical claims are an agent's ASR, the difference in ASR between two agents, and the effect of a defense, $\Delta=\asr_{\text{defended}}-\asr_{\text{undefended}}$ (negative if the defense lowers the ASR). An LLM gate is a decision maker of the same kind: for each proposed action it outputs ALLOW, ASK (escalate to the user), or BLOCK, and a claim about a gate aggregates these decisions.

\subsection{Content-free state}
The \emph{content} of a case is everything a claim is about: the task, the tools and their descriptions, the injected instruction, the policy, and the evidence. A run also depends on inputs that are not content. We call an input \emph{content-free} for a claim if the claim does not mention it and it carries no information about the agent, the attack, or the defense under test: any of its values can occur with the same content, and ``the ASR of agent $X$ under attack $Y$'' says nothing about the interpreter's hash seed. The claim decides what is content: inputs that it does not distinguish are equivalent, and the evaluator declares this equivalence before the runs. A claim about an agent with a set of tools does not distinguish the set's orders, so the rule that orders it is state, even when it sorts by content: the rule carries no information about the attack, although the order that it produces can (Section~\ref{sec:order}); a claim about a deployment that fixes an order makes that order content. Table~\ref{tab:state} lists the state that we study.

\begin{table}[t]
\caption{Content-free state studied in this paper: where it enters the pipeline, who sets it, and how \tare{} handles it.}\label{tab:state}
\centering\footnotesize\setlength{\tabcolsep}{5pt}
\begin{tabularx}{\textwidth}{@{}lYlY@{}}
\toprule
State & Where it enters & Set by & \tare{}\\
\midrule
Hash seed & iteration order of sets (ASB's merge of the attacker tool) & interpreter or evaluator & keyed order; pin and record\\
Ordering rule & sorting of tool lists & harness author & keyed order, averaged over keys\\
Evidence order & member order of a gate's evidence & the requester & keyed presentations that must agree\\
Serving instance & kernels, batching, tensor-parallel layout & platform & same-state rerun (the floor)\\
Failure mapping & verdict for errors and rejected requests & scorer & third verdict\\
\bottomrule
\end{tabularx}
\end{table}

\subsection{Threat model}\label{sec:threat}
\emph{Asset.} The asset is the correctness of a security conclusion: an ASR, a defense effect, a ranking of agents, or a gate decision.

\emph{Adversaries.} Two parties gain from a particular verdict and can choose content-free state. A \emph{strategic evaluator}, for example the author of a defense or of an agent, wants a favorable result and controls the evaluation: the seed, the order in which the harness lists items, and which of several runs is reported. It also gains from a defense that makes the model server reject requests, because a scorer with only two verdicts counts every rejected request as a stopped attack (Section~\ref{sec:failed}). A \emph{strategic requester} of an LLM gate wants a proposed action allowed and controls how the request is assembled, in particular the order of its evidence. In a third, non-adversarial case \emph{nobody} chooses: an unset seed is drawn at random and a sort order is inherited from a library default. Our measurements show that this case alone suffices to change conclusions; the two adversaries can only do better.

\emph{Capabilities.} A party can choose any value of the state that it sets (Table~\ref{tab:state}), and nothing else: not the content of a case, the scorer's success criterion, the model weights, or the serving stack. A requester in our model reorders the evidence but does not add, remove, or reword any of it.

\emph{Trust and the security property.} We trust the benchmark's content and the scorer's success criterion as written (in ASB, whether the attacker's goal appears in the transcript). We do not trust the content-free state, and the scorer's rule for runs that produced no measurement belongs to it: mapping an API error to ``attack failed'' is a choice that the claim does not mention. The security property, decision noninterference (Section~\ref{sec:dni}), requires that the state does not influence the verdict. Because the state can be selected and not merely drawn, a statistical statement about random draws is not enough: \tare{} removes the state where it can, which leaves only a key that the evaluator takes from a public randomness beacon or commits to before the runs, by publishing its hash together with the claim, the declared equivalence, and the analysis, and revealing the key with the results (Section~\ref{sec:strip}), and for the rest it assumes the worst about the direction of the measured changes (Section~\ref{sec:attr}). The two adversaries differ in what \tare{} can do about them. Against the requester of a gate, a canonical order or the stripped gate enforces the property (Section~\ref{sec:procedure}). Against an evaluator, \tare{} is a reporting standard: it cannot stop misreporting, but it makes the choice of state checkable. It removes the choices that can be removed and records the others: a card lists the seeds, the keys, and a hash of the harness code, so that others can repeat a comparison under fresh keys.

\emph{Scope.} We sample the state rather than search it, so our influences are lower bounds on what a selected state can do; content, scorers, weights, and servers are trusted.

\section{Decision Noninterference and \tare{}}\label{sec:tare}
The introduction asked when a reported effect can be attributed to the agent, the attack, or the defense rather than to the state of the run. This section answers in two parts. We first state what a verdict should satisfy, a requirement that we call decision noninterference, and how to measure how far a system is from it (Section~\ref{sec:dni}). We then describe what an evaluator can do when the requirement fails: test whether an effect is larger than anything the measured state could have produced (Section~\ref{sec:attr}), remove the state where it can be removed (Section~\ref{sec:strip}), and test small effects by averaging over state that the evaluator draws (Section~\ref{sec:randomization}). Section~\ref{sec:procedure} applies the same ideas to LLM gates and to failed runs. The formal statements are elementary on purpose: each states what a test guarantees and under which assumption, and several of them fail in current practice (Section~\ref{sec:planted}).

\subsection{Decision noninterference}\label{sec:dni}
Consider one test case of ASB, run twice with nothing changed except the hash seed. The agent, the attack, and the defense are the same in both runs, so a claim about any of them predicts the same verdict twice. If the verdicts differ, the seed caused the difference, and every claim that counts this case is exposed to it. Decision noninterference turns this observation into a requirement.

We write $y_X(c,n)$ for the verdict of system $X$ (an agent, possibly with a defense, or a judge) on case $c$ in a run with content-free state $n$. The state collects the variables that the evaluator declares as content-free, such as the hash seed, the tool order, and the serving instance (Table~\ref{tab:state}); $N$ is the set of their possible values, and every statement below is relative to this declared space. A run can fail to measure anything, as when a request is rejected. We then write $y_X(c,n)=\bot$ and call the case \emph{inconclusive} in that run.

\begin{definition}[Decision noninterference]
System $X$ satisfies decision noninterference (DNI) on cases $C$ with respect to state space $N$ if, for every case $c$ and all states $n,n'$, the verdicts $y_X(c,n)$ and $y_X(c,n')$ have the same distribution.
\end{definition}

In words, changing a detail that the claim does not mention must not change what the benchmark says about any case. The content-free state plays the role of the low-integrity input of noninterference, and the verdict that of the high-integrity output. On a deterministic server the two verdicts must be equal, so a single case whose verdict changes with the state violates DNI. As for noninterference, no single run can show such a violation, but a pair of runs with different states can: on a deterministic server, DNI is a 2-safety hyperproperty~\cite{clarkson2010hyper}. On a stochastic server, where even identical runs can differ, DNI asks only for equal distributions, and a violation must be shown statistically.

DNI holds or fails, but an evaluator also needs to know how far a system is from it. The \emph{influence} of a change of state from $n$ to $n'$ is the share of cases whose verdict differs between the two runs,
\[ I_X(n,n')=\frac{1}{|C'|}\sum_{c\in C'} \mathbf{1}[y_X(c,n)\neq y_X(c,n')], \]
where $C'$ is the set of cases that are conclusive in all runs of the comparison. The \emph{floor} is the influence between two runs with the same state, that is, of a plain rerun. It is zero on a deterministic server and positive on a stochastic one, and an influence carries information about the state only where it exceeds the floor. An excess beyond same-state variation shows a violation of DNI, but its absence does not show DNI: a case that succeeds with probability 0.5 under one state and 0.6 under another disagrees across them as often as across two runs of the first. On a stochastic server, DNI is thus tested on outcome frequencies from repeated runs (Section~\ref{sec:randomization}). Cases that are conclusive in one run and inconclusive in the other have also changed their outcome, and \tare{} reports their share separately. Finally, an influence of zero over the values that were tried does not establish DNI for the values that were not; Section~\ref{sec:strip} shows how DNI can instead be obtained by construction.

\subsection{The attribution test}\label{sec:attr}
Influence describes single verdicts, but claims are about aggregates. A typical claim compares two systems $A$ and $B$, for example a defended and an undefended agent, by the difference of their attack success rates, $\Delta=\asr_A-\asr_B$. The two runs may have different states, $n_A$ and $n_B$, so we write $\Delta(n_A,n_B)$. Attribution asks how much this difference could change if the states were different.

\begin{proposition}[Sensitivity bound]\label{prop:bound}
For any states $n'_A$ and $n'_B$, with all quantities computed on the cases $C'$ that are conclusive in the four runs,
\[ |\Delta(n_A,n_B)-\Delta(n'_A,n'_B)|\le I_A(n_A,n'_A)+I_B(n_B,n'_B), \]
with equality if and only if all changed verdicts move $\Delta$ in the same direction.
\end{proposition}
\begin{proof}
Each case whose verdict changes moves an ASR by one case, that is, by $1/|C'|$. Hence the ASR of $A$ moves by at most the share of $A$'s verdicts that change, $I_A$, and likewise for $B$. The difference moves by at most the sum, and by exactly the sum if no two changes cancel.
\end{proof}

The bound is elementary, but it has a useful consequence: an effect that is larger than the sum of the two influences cannot change its sign under any change of state of that size.

\begin{corollary}[Attribution]\label{cor:attr}
Let $|\Delta(n_A,n_B)|>m_A+m_B$. Then, on the same cases, $\Delta$ keeps its sign at all states $n''_A$, $n''_B$ that change at most a share $m_A$ of $A$'s verdicts and $m_B$ of $B$'s, that is, with $I_A(n_A,n''_A)\le m_A$ and $I_B(n_B,n''_B)\le m_B$, whatever the direction of the changes.
\end{corollary}

\tare{} turns the corollary into a test. It runs each system at a second state, fixed in advance or drawn at random so that no party can pick one with a small influence, measures the two influences, floor included, and uses their sum as the \emph{margin} $m_A+m_B$, a worst case rather than a standard error. It \emph{attributes} the effect to $A$ versus $B$ only if both measured effects, $\Delta(n_A,n_B)$ and $\Delta(n'_A,n'_B)$, exceed the margin in absolute value. Otherwise it reports the effect as \emph{not attributable at this margin}, which is not the same as ``no effect''. In the example of the introduction, a change of seed changes 59 of the undefended and 44 of the defended agent's 400 verdicts. The margin is therefore $(59+44)/400=25.8$ points, far above the measured drops of 4.5 and 1.8 points.

Two remarks delimit what an attribution establishes. First, it rules out the content-free state as the explanation of the effect's sign, for every state that moves no more verdicts than the measured change; it does not prove that the defense causes the effect. It also presupposes that the two arms differ in nothing else, so every state variable must be equal in both arms or varied within each arm, as the seed is here. \tare{} therefore records the declared state of each run, including a hash of the harness code, and checks that the runs of a comparison share it.

Second, the test is conservative on purpose: the margin counts every changed verdict as if it moved against the effect. A statistical test would instead compare the effect with the typical spread of the ASR under randomly drawn states (Section~\ref{sec:randomization}). That comparison fits state that the evaluator draws at random, but not state that a party can select, nor state that pushes all cases in one direction. A sort order does the latter: under sorting, 84 of the 93 changed verdicts of Llama-3.1-8B move the same way, and the ASR rises by 18.8 points (Section~\ref{sec:order}). The full margin protects against an unmeasured state that changes as many verdicts as the measured change and changes every one of them against the effect; if each case merely takes one of its two measured verdicts, half the margin suffices (Appendix~\ref{app:guarantees}). The guarantee does not extend to states that move more verdicts, and such states exist: over three to six seeds, 1.5 to 2.3 times as many verdicts depend on the seed as differ between two (Section~\ref{sec:seed}). A margin from two values of one variable is a lower bound on what that variable can do and says nothing about variables that were not varied. The test needs only two runs per system and assumes nothing about how the state was chosen.

\subsection{Removing content-free state}\label{sec:strip}
The attribution test measures the state; where possible, the better remedy is to remove it. How depends on the kind of state.

\emph{Process state}, for example \texttt{PYTHONHASHSEED}, is pinned and recorded. Pinning makes reruns identical, but it selects one value of the state, so on its own it hides influence rather than removing it.

\emph{Presentation state}, the order in which tools or evidence are shown to the model, can be removed. The idea is to compute the order from a value that no other party controls and that no claim mentions: a key that the evaluator draws.
\begin{definition}[Keyed order]\label{def:keyed}
Let $k$ be a 256-bit evaluator key, $c$ a case identifier, and $S$ an unordered collection declared by the evaluator (the tools of a plan, the members of an evidence object). The keyed order of $S$ lists its items by increasing $\mathrm{HMAC\text{-}SHA256}(k_c,\mathrm{canon}(i))$, where $k_c=\mathrm{HMAC\text{-}SHA256}(k,c)$ is the per-case key and $\mathrm{canon}(i)$ is a canonical serialization of item $i$.
\end{definition}
In words, each case gets its own shuffle of its items, determined only by the key and by the items' content. The shuffle ignores the order in which the items arrived and the interpreter's hash seed. If these reach the model only through the collection, a harness with a keyed order therefore satisfies DNI with respect to all arrival orders and seeds, by construction rather than by sampling. Names enter only through the keyed hash, so no naming convention can put one kind of tool first, as sorting by name does (Section~\ref{sec:order}). The verdict still depends on the key, because some order must be shown. The difference is that the evaluator draws the key at random and records it: no other party selects it, and its effect can be averaged and tested. For an offline benchmark, a keyed order is simply a reproducible random permutation per case; the key must stay secret only when the party that supplies the items must not predict their positions, as for the requester of a gate.

\emph{Averaging over keys} then replaces the single run. With $K$ runs under independent keys, the averaged verdict of a case is its success frequency $f_X(c)=\frac{1}{K}\sum_{j=1}^{K}y_X(c,n_j)$, and the averaged ASR, $\asr_K$, is the mean of these frequencies. If the keyed hash behaves like a random function, a random key gives each case a uniformly random order, so $\asr_K$ is an unbiased estimate of the ASR under uniformly random positions of the items, unlike the ASR under any fixed order (Section~\ref{sec:order}); we call an evaluation with \tare{} unbiased in this sense only. How precise is the estimate? Let $p_c$ be the probability that case $c$ succeeds under a random key.

\begin{proposition}[Variance of the averaged ASR]\label{prop:var}
If the keys are independent, two runs disagree on an expected share
\[ \bar I=\frac{2}{|C|}\sum_{c\in C} p_c(1-p_c) \]
of the cases. If, in addition, the verdicts of different cases vary independently, the averaged ASR has variance
\[ \mathrm{Var}\big(\asr_K\big)=\frac{1}{|C|^{2}K}\sum_{c\in C}p_c(1-p_c)=\frac{\bar I}{2\,|C|\,K}. \]
\end{proposition}

The proof is a short calculation (Appendix~\ref{app:guarantees}). Its practical value is that the noise that the state adds to an ASR follows from a quantity that is easy to measure, the share of cases on which two runs disagree. For example, two keys disagree on 13.6\% of Llama-3.1-8B's 400 cases (Section~\ref{sec:kdraw}), so the ASR of a single run has a standard deviation of about 1.3 points, and an average over eight keys one of about 0.5 points. The formula assumes that cases vary independently. A global hash seed breaks this assumption, because one seed orders every set in the process at once; per-case keys give each case its own draw, which alone decides its verdict on a server that runs one request at a time (Section~\ref{sec:planted}).

\subsection{Testing averaged comparisons}\label{sec:randomization}
Averaging makes small effects visible, but deciding whether an averaged effect is real requires a test, and \tare{} offers two kinds. The \emph{margin test} keeps the worst-case guarantee of Section~\ref{sec:attr}: it splits each system's $K$ runs into two halves and applies Corollary~\ref{cor:attr} to the success frequencies of the halves (Appendix~\ref{app:guarantees}). It, too, needs large effects.

The \emph{randomization test} instead uses the fact that the evaluator draws the keys at random. Suppose that the defense changes nothing. Then a defended run is just another draw of the undefended system, and the labels ``defended'' and ``undefended'' carry no information about the runs. The test therefore compares the observed difference in mean ASR with the differences obtained when the labels are reassigned in every possible way, $K$ runs to each arm; its p-value is the share of reassignments whose difference is at least as large, in absolute value, as the observed one. With $K=6$ there are 924 reassignments, so the smallest possible p-value is $2/924\approx0.0022$ (0.029 with $K=4$). The test is exact if each run draws its own key, independently of any outcome (a key list committed before the runs makes this checkable), and if the runs are otherwise executed alike, with the same harness code and on one deterministic serving instance or on interchangeable ones (which a same-key repeat checks). It needs no model of how cases vary, in particular no independence between them; it is Fisher's randomization inference~\cite{fisher1935design,imbens2015causal} applied to evaluation runs.

Because the randomization test compares runs under different keys, a chance difference between the keys of the two arms can look like an effect. \emph{Matched keys} avoid this: both arms run under the same $K$ keys, so every case has the same order in both arms, and the main effect of the key cancels in the $K$ per-key differences of the ASR. An interaction between key and defense, that is, a defense that works better under some orders than under others, remains part of the effect. The \emph{matched test} flips the signs of the $K$ differences in all $2^K$ ways and reports the share of sign patterns whose sum is at least as large, in absolute value, as the observed one; its smallest p-value is $2/2^K$. Its null hypothesis is that the differences are independent and each is distributed symmetrically about zero.

The two kinds of test answer different questions. In this paper an effect is \emph{attributed} only by the margin test, which applies even to runs whose state the evaluator did not draw, but needs large effects. The randomization and matched tests \emph{detect} small effects at a stated error rate, but only in runs whose keys the evaluator drew. In practice, the margin test suits state that cannot be stripped or may have been selected, such as a fixed order or a stochastic server; once the order is keyed and the keys are committed, the matched test is the one to use for small effects. Table~\ref{tab:guarantees} in the appendix lists what each part of \tare{} guarantees.

\subsection{Gates, failed runs, and the procedure}\label{sec:procedure}
\emph{Gates.} An LLM gate faces the same state as a benchmark, with one difference: the party that submits a request also chooses the order of its evidence. \tare{} therefore applies keyed presentations online, and we call the result a \emph{stripped gate}. For each request, the stripped gate builds $K$ presentations of the evidence under independent keys, asks the judge about each, and returns a decision only if at least a share $\tau$ of the $K$ answers agree on it. Otherwise it abstains, and the abstention is handled fail-closed: the action is escalated to a human or denied. The gate draws its keys once and keeps them secret: with fresh keys for every submission, a requester could resubmit a request until some presentation is allowed, whereas with fixed keys a resubmission meets the same presentations. For the same reason, it derives the case identifier of Definition~\ref{def:keyed} from the request's canonical serialization, not from an identifier that the requester supplies.

\begin{proposition}[Order independence of the stripped gate]\label{prop:gate}
For a deterministic judge, the stripped decision depends only on the content of the request and on the evaluator's keys; in particular, it is the same for every order in which the request's unordered parts are submitted.
\end{proposition}
\begin{proof}
A keyed order depends only on the key and the items' content, so all submitted orders of the same items produce the same $K$ presentations, answers, and decision.
\end{proof}

The proposition does not make an order-sensitive judge reliable; it moves the dependence on order from the requester to the evaluator's keys. A judge whose answers depend strongly on order will often disagree with itself across keys and abstain, so its abstention rate measures that dependence (Section~\ref{sec:gatestrip}). For a stochastic judge the proposition holds for the distribution of the decision, and the gate should answer a resubmitted request with its first decision.

\emph{Failed runs} receive a third verdict. \tare{} scores a case as inconclusive if its run ended in an API or serving error, a rejected request, a client crash, or truncation before any action, and it reports the inconclusive share next to every ASR. A case in which the agent made no tool call is different: the agent was measured and did not act, so the attack failed. Because a defense can lower the ASR simply by stopping the agent from acting, \tare{} also reports the tool-action rate (the share of cases with at least one tool call) and the ASR among acting cases.

\emph{The procedure} has the five steps of Figure~\ref{fig:pipeline}b, with Table~\ref{tab:state} as the checklist for the first step and a one-page \tare{} card as the report; the implementation is a dependency-free Python package of about 650 lines (Appendix~\ref{app:artifact}).

\section{Experimental Setup}\label{sec:setup}

\emph{Benchmarks.} We study three agent-security benchmarks that a recent defense paper uses together~\cite{bhagwatkar2025firewalls}, AgentDojo, ASB, and InjecAgent, and take the inputs of our gate experiments from $\tau^2$-bench~\cite{tau2}. We analyze ASB in most depth, because its harness derives the order of tools from a Python set. On ASB (commit \texttt{1f561dc}) we use observation prompt injection with the combined attack (400 cases). ASB's three prompt-level defenses are instructional prevention (a warning wrapped around the task), the sandwich defense (the task repeated after every tool result), and delimiters (markers around the task, which the system prompt refers to). The hash seed is set per run through \texttt{PYTHONHASHSEED}; the tool order is ASB's own (set iteration), sorted, or reverse-sorted. On AgentDojo (commit \texttt{089ed46}), a replay without model calls found the 35 of 726 v1 cases in which the seed changes what the agent sees; we test all 35. InjecAgent~\cite{injecagent} (commit \texttt{f19c9f2}), whose tool lists have a fixed order, is a negative control.

\emph{Agents.} We evaluate seven open-weight agents from four families: Llama-3.1-8B-Instruct and Llama-3.1-70B-Instruct (AWQ, 4-bit), Qwen3-8B, Qwen3.5-9B and Qwen3-32B, Gemma-3-27B-it, and Mistral-Nemo-Instruct-2407 (12B), each at a pinned model revision (listed in the artifact). Four further open-weight models, Granite-4.1-8B, OLMo-3-7B-Instruct, Phi-4-mini-instruct, and Qwen2.5-7B-Instruct, produced no tool calls in ASB's format and are not used. Three models are reached through production APIs: Gemini 3.8 Flash and Gemini 4 Argon serve as agents on ASB and AgentDojo, and K2-Horizon-375B, an open-weight mixture-of-experts model served by its developer's API, serves as an agent on AgentDojo and as a judge; Gemini 3.8 Flash is also a judge. Qwen3.8-27B (a different model from Qwen3-8B) serves as a judge.

\emph{Serving.} Open-weight agents are served with vLLM~0.30.0 on RTX 5000 Ada GPUs in a deterministic configuration: one sequence at a time, no prefix caching, no chunked prefill, greedy decoding (temperature 0). Qwen3-32B and Gemma-3-27B (8-bit weights) and Llama-3.1-70B (4-bit) run on two GPUs, the other agents in bfloat16 on one. We verify determinism instead of assuming it: every ASB agent has an exact repeat with identical declared state on the same server instance, and every judge is asked repeatedly about an identical presentation. For two of the three two-GPU agents we also rerun on a second machine.

\emph{Changes to ASB.} Our copies of ASB differ from the pinned commit only in ways that leave the content of every message unchanged (Appendix~\ref{app:artifact}). \tare{}'s check of the harness code of all 229 runs confirms that the compared runs of an agent share the same code, with four exceptions that Section~\ref{sec:failed} reports. For the keyed-order experiments of Section~\ref{sec:controlled}, a short patch makes ASB's merge of the attacker tool use the keyed order of Section~\ref{sec:strip} when an evaluator key is set; without a key the code is unchanged. All agents use one list of evaluator keys.

\emph{Judges.} We take 98 consequential tool-call proposals from $\tau^2$-bench conversations without selecting on outcome. Each proposal is a JSON evidence packet with the policy, the conversation, and the proposed call, and a judge returns ALLOW, ASK, or BLOCK. Each packet is presented in its canonical member order, in ten (eight for K2-Horizon-375B) independent random member orders, and in repeated canonical presentations (Table~\ref{tab:judges}); every byte of content is the same in all presentations. The judges of the main study are Qwen3.8-27B, K2-Horizon-375B, Gemma-3-27B, Qwen3.5-9B, Llama-3.1-8B, and Mistral-Nemo-12B. We later added Gemini 3.8 Flash as a seventh judge; because of its cost it sees six random orders and one repeat, a design that we fixed before its runs.

\emph{Statistics.} An influence is the share of cases with a changed verdict, averaged over the pairs of distinct seeds. Order effects are tested with exact McNemar tests on the cases that change between two runs, with Holm's correction for multiple tests. These tests treat the cases as independent. We therefore add a sign-flip test over ASB's ten scenarios, which needs no independence between the cases of a scenario: it compares the sum of the ten per-scenario differences with its distribution when the sign of each difference is flipped at random (null: independent differences, symmetric about zero). Inconclusive cases are excluded and counted.

\section{Content-Free State Reaches Verdicts}\label{sec:findings}
\begin{figure*}[t]
\centering
\includegraphics[width=\textwidth]{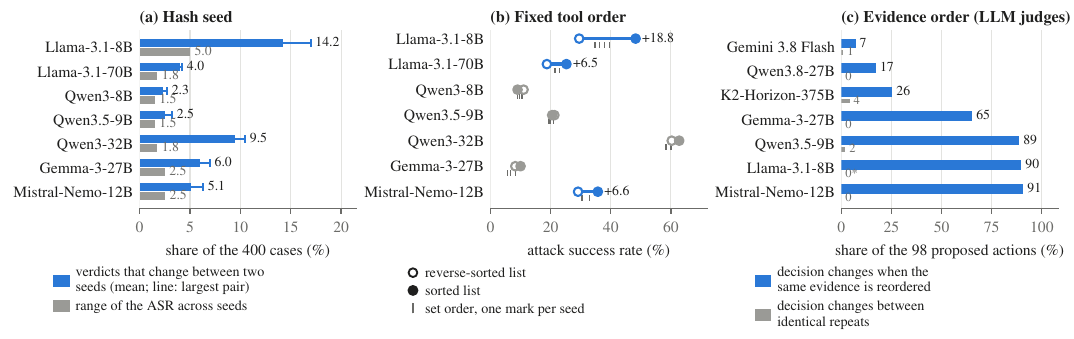}
\caption{Content-free state changes verdicts while everything a claim is about stays fixed. (a)~ASB: two seeds differ in 2.3--14.2\% of the 400 verdicts on average and in up to 17.0\%, while the ASR moves much less; an identical rerun changes none. (b)~The choice between two fixed tool orders moves the ASR of three agents (blue) by 6.5 to 18.8 points. (c)~Proposals on which a judge's decision differs across orders of the same evidence (blue; eleven orders, nine for K2-Horizon-375B, seven for Gemini 3.8 Flash) and across identical repeats (gray).}\label{fig:effects}
\end{figure*}

We follow content-free state through the pipeline of Figure~\ref{fig:pipeline}b, from the state that nobody chooses (the hash seed, Section~\ref{sec:seed}) to choices made once in harness code (a fixed tool order, Section~\ref{sec:order}), state that a requester controls (the order of a judge's evidence, Section~\ref{sec:judges}), and the scorer's rule for failed runs (Section~\ref{sec:failed}); Section~\ref{sec:defenses} then asks what they do to conclusions about defenses. Figure~\ref{fig:effects} summarizes the largest dependences; Appendix~\ref{app:tables} has all tables.

\subsection{The hash seed decides ASB verdicts}\label{sec:seed}
ASB's verdicts are reproducible, yet they depend on the hash seed (Figure~\ref{fig:effects}a; Table~\ref{tab:seed} in the appendix). For every agent, two runs with the same hash seed on the same server instance give the same outcome on every case. Runs with different seeds do not. Between two seeds, Llama-3.1-8B changes 14.2\% of its verdicts on average and up to 17.0\%; the other six agents change 2.3--9.5\% on average. The ASR hides most of this, because changes in both directions cancel: the ASR of Llama-3.1-8B spans 5.0 points across five seeds, and Gemma-3-27B changes 7.0\% of its verdicts between two seeds at an identical ASR of 8.2\%. More seeds reveal more dependence: the share of cases whose verdict differs between any two of an agent's three to six seeds is 1.5 to 2.3 times the share that differs between two seeds.

The cause is one function of the harness. ASB inserts the attacker tool into the agent's plan by collecting the names of the plan's tools and of the attacker tool in a Python \texttt{set} and assigning \texttt{list(set)} to every stage of the plan. String hashing in Python is randomized per process~\cite{pythonhashseed}, a defense against hash-flooding attacks~\cite{crosby2003,pep456}, so the iteration order of the set, and with it the position of the attacker tool in the list that the agent sees, depends on \texttt{PYTHONHASHSEED}. ASB does not set this variable, so each fresh process draws its own seed. In our runs the tool order is the only path from the seed to the verdict. Of 20,770 comparisons of one case under two seeds (all seed pairs of the seven agents, one machine per agent), the 5,439 in which the two seeds happened to give the case the same tool order never differ in verdict; all 1,313 differing verdicts occur among the other 15,331. Two controls agree. InjecAgent lists tools in a fixed order, and a change of seed there changes none of 397 verdicts. On AgentDojo, where the seed enters through the order of simulated participants, Llama-3.1-8B changes 6 of the 35 cases between seeds and none between repeats.

The serving instance is state of the same kind. Every repeat within an instance is identical, but two instances with identical flags can differ, in up to 116 of 397 verdicts at the same seed (Appendix~\ref{app:tables}). We therefore compare seeds, orders, and defenses within one instance, or across two after an exact same-state control.

A stochastic server can hide the seed altogether, because the seed's influence mixes with the server's own variation. Through production APIs, identical runs change 13.8\% and 8.1\% of the ASB verdicts of Gemini 3.8 Flash and Gemini 4 Argon and 10 of 35 AgentDojo outcomes of K2-Horizon-375B (none of either Gemini model's), and the influence of the seed cannot be separated from these floors, so Figure~\ref{fig:effects}a omits these agents. Only where a same-state rerun is identical can a changed verdict be traced to the harness; there, a benchmark can be reproducible yet depend on a value that nobody chose.

\subsection{A fixed tool order decides the ASR}\label{sec:order}
A natural fix for an order that varies is an order that does not, for example a sorted tool list. Such a fix makes reruns agree, but it moves the ASR (Figure~\ref{fig:effects}b; Table~\ref{tab:order}). Between a reverse-sorted and a sorted list at a fixed seed, the ASR of three of the seven agents rises by 6.5 to 18.8 points, and for the other four it changes by at most 2.5 points. Llama-3.1-8B rises from 118 to 193 successful attacks of 400: 84 cases succeed only when sorted and 9 only when reverse-sorted. The increase is significant for all three agents with the case as the unit, and for the two Llama agents also with the scenario as the unit (Table~\ref{tab:order}). The choice also changes comparisons between agents. Under the reverse-sorted order Llama-3.1-8B and Mistral-Nemo-12B are 0.4 points apart; under the sorted order Llama-3.1-8B appears 12.6 points more vulnerable, and two other pairs of agents swap ranks.

The direction of the effect comes from the benchmark. All 400 attacker tools in ASB (40 per scenario) have CamelCase names (for example, \texttt{DiagnosisManipulation}), while all 20 benign tools use snake\_case. In ASCII, uppercase letters precede lowercase ones, so sorting places the attacker tool first and reverse sorting places it last. The size of the effect depends on the agent: agents that favor the first listed tool, a bias documented for tool selection~\cite{biasbusters2025,faghih2025tool}, call the attacker tool more often when it comes first.

The names are content, but the rule that orders by them is state: an order computed from names inherits the naming habits of whoever wrote the tools. Averaging over keyed orders instead estimates the ASR over uniformly random positions, 38.1\% for Llama-3.1-8B, between the two fixed orders (Section~\ref{sec:kdraw}). Which summary is right depends on the claim. If an attacker can choose where its tool is listed, the position belongs to the content of the claim, and the relevant figure is the ASR at the worst position (48.2\% here). \tare{} requires only that the choice be declared.

\subsection{Evidence order decides LLM-gate decisions}\label{sec:judges}
LLM judges that authorize actions change their decision when the same evidence is reordered (Figure~\ref{fig:effects}c; Table~\ref{tab:judges}). We present each of 98 unselected proposals to the six judges of the main study in its canonical member order and in ten random orders (eight for K2-Horizon-375B). Counted per proposal, a judge gives at least two different decisions among these presentations for 17\% of the proposals for Qwen3.8-27B, 26\% for K2-Horizon-375B, 65\% for Gemma-3-27B, and 89--91\% for Qwen3.5-9B, Llama-3.1-8B, and Mistral-Nemo-12B; among repeats of one presentation it does so for at most 4 of 97 proposals. Counted per pair of presentations, these judges decide differently on two random orders in 7--47\% of the pairs (Table~\ref{tab:gate}, $K=1$) and on two identical presentations in 0--2.8\%. The changes include the direction that matters for a gate: 71--86\% of the proposals not allowed under the canonical order are allowed under at least one random order.

The split decisions are not confined to borderline cases. For Qwen3.8-27B and K2-Horizon-375B, the two most stable of the six, each of the 17 and 25 proposals with split decisions is allowed under one order and blocked under another, as are 70 of the 87 for Qwen3.5-9B; for the other three judges, 82--92\% of the splits are between ALLOW and ASK only. Gemini 3.8 Flash, the seventh judge, is the most stable: its decision differs across orders for 7 of the 98 proposals (1 of 98 between identical presentations), all between ALLOW and BLOCK, whereas on the same seven orders the other judges split on 13--85\%. For the most stable judges, the dependence is thus of the most consequential kind: the order of the evidence, which the requester chooses, can decide whether an action is allowed.

\subsection{Failed runs become verdicts}\label{sec:failed}
ASB has two verdicts and no third value, so every run that fails for reasons unrelated to security is scored as an attack that failed (Table~\ref{tab:failed} in the appendix). ASB's client sends two leading system messages, which the chat templates of Gemma-3-27B and Qwen3.5-9B reject. In four runs with the unmodified client, two per model, every one of 400 requests was rejected, and ASB reported an ASR of 0\% for each run, although no request reached a model. These runs were meant as our delimiters arms; we had launched them by mistake without the client change of Section~\ref{sec:setup}, and we reran them with it (Appendix~\ref{app:n15}). A defense can produce the same failure in part. Under the delimiters defense, Mistral-Nemo-12B's chat template rejects 64--71 of 400 cases, partly different ones at the two seeds, against 4 without the defense. ASB scores them as failed attacks and reports an ASR of 30.0\%, about what it reports for the undefended agent (30.5\%).

The failures have ordinary causes; the defect is that the verdict cannot express them, so that a broken run counts as a secure one. \tare{}'s third verdict marks these cases inconclusive when the transcript records the failure. The error-scored runs then change from ``secure'' to ``not measured'', and Mistral-Nemo-12B's delimiters defense no longer looks neutral: on the 320 cases that reached the model in all four runs, it raises the ASR from 30.3\% to 35.6\%. Because a comparison on conclusive cases is conditional on them, \tare{} reports the inconclusive share with every ASR.

\subsection{Defense effects and what they consist of}\label{sec:defenses}
\begin{figure}[t]
\centering
\includegraphics[width=0.62\textwidth]{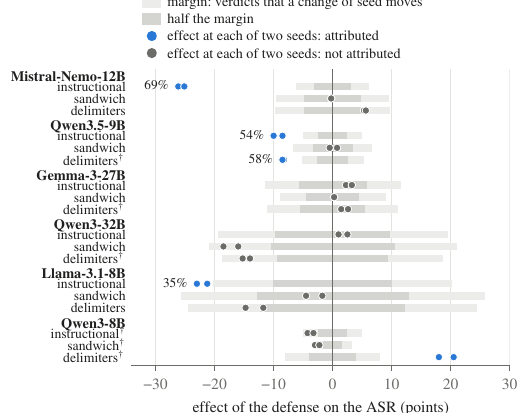}
\caption{The margin test on ASB's defenses, eighteen combinations of agent and defense: effects at two seeds (dots) against the margin $m_A+m_B$ (gray band). Attributed (blue): both dots outside the band. A mean outside the darker inner band: no case-by-case mix of the two seeds flips the sign. \%: stopped attacks without a tool call, for the attributed decreases. $^\dagger$Run afterwards under a design fixed in advance (Appendix~\ref{app:n15}).}\label{fig:defense}
\end{figure}

The previous sections showed that content-free state changes verdicts; we now ask whether it changes conclusions. We apply the attribution test to ASB's three defenses on six of the seven agents, eighteen combinations of agent and defense, each measured at two seeds (Figure~\ref{fig:defense}; Table~\ref{tab:defenses}). The margins are with respect to the hash seed, the only state that ASB's own harness varies. Twelve combinations were measured first; the other six ($\dagger$) were run afterwards under a design fixed in advance, each on a new serving instance with an undefended control as its base arm (Appendix~\ref{app:n15}). We did not run Llama-3.1-70B, whose runs need about two thousand model calls each on two GPUs.

\emph{Only large effects can be attributed.} The margins range from 3.2 to 25.8 points, so on most agents only a large effect can pass the test; five effects of two defenses do. Instructional prevention is attributed on three of the six agents: it lowers Mistral-Nemo-12B's ASR by 26.2 and 25.1 points at the two seeds against a margin of 6.2 points, Llama-3.1-8B's by 23.0 and 21.2 against 20.2, and Qwen3.5-9B's by 10.0 and 8.5 against 5.0. Delimiters are attributed on two agents, in opposite directions: they lower Qwen3.5-9B's ASR by 8.2 and 8.5 points against a margin of 5.2, and raise Qwen3-8B's by 18.0 and 20.5 against 8.0. On the other four agents they are not attributed; on Llama-3.1-8B, for example, they lower the ASR by 14.8 and 11.8 points against a margin of 24.5. All five attributions survive the worst treatment of the cases excluded as inconclusive. With a margin from the largest influence between any two of the undefended agent's seeds, four remain, and the one on Llama-3.1-8B does not (22.5 against 21.2 points).

\emph{A clear effect can still fail the test.} The sandwich defense is never attributed, and its effect on Qwen3-32B shows why. The defense lowers this agent's ASR by 18.5 and 16.0 points, many times the 1.8-point range of its ASR across four seeds. But a change of seed moves about 10\% of this agent's verdicts with and without the defense, a margin of 21.0 points, so a state that moved as many verdicts, all in one direction, could produce an effect of this size. Still, the mean of its two values exceeds half the margin in absolute value. Its sign therefore survives every per-case combination of the two measured seeds, as do the signs of five other effects that are not attributed (marked $^\circ$ in Table~\ref{tab:defenses}; for delimiters on Mistral-Nemo-12B, on the 320 cases that all four runs measured). These effects fail only the full margin, which guards against unmeasured states; Section~\ref{sec:defense-keyed} returns to the difference between an attributed and a detected effect.

\emph{An attributed drop is not yet a safer agent.} Attribution says that a drop in ASR is not explained by the measured content-free state. It does not say that the agent became safer, and for both defenses with an attributed decrease, much of the drop is the agent no longer acting. On Mistral-Nemo-12B, the agent with the largest attributed effect, the tool-action rate (seed 0, 390 conclusive cases) falls from 94.6\% to 40.8\% under instructional prevention, so the defended agent makes no tool call in 59.2\% of these cases. Of the 110 attacks that succeed on the undefended agent but fail under the defense, 76 (69\%) end without any tool call by the defended agent, which answers each step with a text plan instead of acting, and ASB's scorer counts each as a thwarted attack. For Qwen3.5-9B the share is 54\% and for Llama-3.1-8B 35\%. The effect is not only inaction. Among the cases in which the agent still acts, the ASR falls as well, for Mistral-Nemo-12B from 32.8\% to 11.9\%. The defense also costs the benign task: ASB's task success falls from 50.5\% to 17.0\% for Llama-3.1-8B and from 33.5\% to 18.0\% for Qwen3.5-9B (Mistral-Nemo-12B completes 2.3\% of its tasks with and without it).

\emph{Delimiters change how often agents act.} The attributed decrease from delimiters on Qwen3.5-9B shows the pattern without any gain among acting cases: the agent calls a tool in 44.2\% of the cases instead of 80.8\% (seed~0), 32 of the 55 attacks that the defense stops (58\%) end without any tool call, and among the cases in which it acts, the ASR does not fall (26.0\% instead of 24.5\%; at seed~1, 24.4\% instead of 24.6\%). The attributed increase on Qwen3-8B goes the other way: under delimiters, this agent calls a tool in 82.5\% of the cases instead of 50.2\%, and when it acts, the attack succeeds more often (33.6\% instead of 19.4\%). Instructional prevention, below its margin on Qwen3-8B, again stops attacks mostly by stopping the agent from acting (27 of 30). A lower ASR can thus come from a safer agent, from an inactive one, or from the state of the run, and only the first is a security gain; telling them apart takes the attribution test, the action rate, and the ASR among acting cases.

\section{Controlled Experiments: What \tare{} Changes}\label{sec:controlled}
\begin{figure*}[t]
\centering
\includegraphics[width=\textwidth]{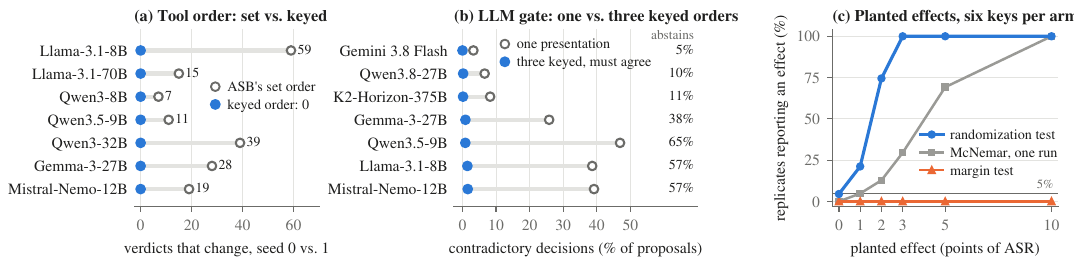}
\caption{What \tare{} changes. (a)~A change of hash seed changes 7 to 59 of 400 verdicts under ASB's set order and none under the keyed order. (b)~Proposals on which two independent evaluations contradict each other, with one presentation (hollow) or three keyed presentations that must agree (filled); right: abstentions. (c)~Planted effects of known size, Llama-3.1-8B, six keys per arm: share of 2,000 replicates in which each test reports an effect.}\label{fig:tare}
\end{figure*}

This section answers RQ3, first where the correct outcome is known (Figure~\ref{fig:tare}), then on real defenses and gates.

\subsection{Keyed orders remove the seed}\label{sec:kdraw}
If the seed reaches the verdict only through the tool order, replacing ASB's set order by a keyed order should make the seed irrelevant. We test this prediction with the keyed-order patch (Section~\ref{sec:setup}). Three agents from three families, Llama-3.1-8B, Mistral-Nemo-12B, and Qwen3.5-9B, run under twelve evaluator keys each, and we repeat their first key with a different hash seed and on other server instances or nodes; the other four agents run the first key at two hash seeds.

\emph{The seed no longer matters} (Figure~\ref{fig:tare}a; Table~\ref{tab:keyed} in the appendix). Under a fixed key, changing the seed from 0 to 1 changes no verdict of any of the seven agents, whereas under ASB's set order it changes between 7 and 59 of their 400 verdicts. For the three agents with twelve keys, the same key also gives identical verdicts on another node four days later, and for two of them on other instances.

\emph{Keys still change verdicts, but the evaluator draws them.} Some order must be shown, so a change of key changes verdicts much as a change of seed does: 13.6\% of Llama-3.1-8B's verdicts on average between two keys, against 14.2\% between two seeds. The difference lies in who chooses. The evaluator draws the key and records it, so its effect can be averaged: the mean ASR over twelve keys, 38.1\%, lies between the sorted (48.2\%) and the reverse-sorted order (29.5\%), as an average over random positions should.

\subsection{Calibration and power}\label{sec:planted}
A test is useful only if it reports no effect where there is none and finds the effects that exist. We check both with comparisons whose truth we know.

\emph{Null pairs.} Two runs of the same undefended agent at different seeds compare a system with itself, so every difference between them is noise. Proposition~\ref{prop:var} predicts the size of this noise if cases vary independently. Under a global seed the ASR varies more than predicted: averaged over the seven agents, its variance is 1.7 times the prediction, a ratio that a simulation of independent cases produces in only 1.6\% of 20,000 replicates. Under per-case keys the ratio is 0.9. A test that treats cases as independent is therefore overconfident under a global seed, but not under keys (Appendix~\ref{app:planted}).

\emph{Planted effects.} To measure power, we plant synthetic defenses into the keyed runs: a replicate assigns six of an agent's twelve runs at random to each arm and makes a random set of cases fail in every run of one arm (Figure~\ref{fig:tare}c; Appendix~\ref{app:planted}). For Llama-3.1-8B, the agent with the largest influence, the randomization test finds a planted effect of 3 points in every replicate and one of 2 points in 74.5\% of them, whereas a McNemar test on one run per arm finds them in 29.5\% and 12.8\%. The margin test attributes no planted effect of up to 10 points for this agent; this is the price of its guarantee. These synthetic effects come from one pool of twelve runs, so the rates only illustrate power.

\subsection{Averaged defense comparisons}\label{sec:defense-keyed}
We now apply the averaged tests to real defenses: instructional prevention and the sandwich defense, on the three agents with twelve keys (Table~\ref{tab:n5}). The question is whether averaging can settle what single runs could not, in particular the small sandwich effect of the introduction.

\emph{Design.} We proceeded in three steps. With \emph{disjoint keys}, the defended arm runs under keys 0--5 and the undefended arm under keys 6--11, as the randomization test requires. With \emph{matched keys}, both arms run under keys 0--7 and see the same tool order in every case. Both steps extended earlier runs, so their p-values are exploratory (Appendix~\ref{app:averaged} gives the history). The third step tests the two small effects under eight fresh keys in both arms, with a design that we fixed before any of its runs.

\begin{table}[t]
\caption{Averaged defense comparisons on ASB (effects in ASR points). Keys 0--11 are exploratory: disjoint keys (defended arm 0--5, undefended 6--11; margin and randomization tests) and matched keys (both arms 0--7; matched test). Keys 12--19 were fixed in advance. $p$: Holm-adjusted within each design.}\label{tab:n5}
\centering\footnotesize\setlength{\tabcolsep}{5pt}
\begin{tabular}{@{}lrrcrrr@{}}
\toprule
 & \multicolumn{4}{c}{Disjoint keys} & \multicolumn{2}{c}{Matched keys}\\
\cmidrule(lr){2-5}\cmidrule(l){6-7}
Defense and agent & effect & margin & attributed & $p$ & effect & $p$\\
\midrule
\multicolumn{7}{@{}l}{\emph{Instructional prevention}}\\
\quad Llama-3.1-8B & $-25.0$ & 12.6 & yes & 0.013 & $-26.3$ & 0.047\\
\quad Qwen3.5-9B & $-9.1$ & 3.2 & yes & 0.013 & $-9.7$ & 0.047\\
\quad Mistral-Nemo-12B & $-27.1$ & 4.9 & yes & 0.013 & $-27.3$ & 0.047\\
\multicolumn{7}{@{}l}{\emph{Sandwich}}\\
\quad Llama-3.1-8B & $-3.2$ & 15.6 & no & 0.013 & $-4.6$ & 0.047\\
\quad Qwen3.5-9B & $+1.2$ & 4.0 & no & 0.013 & $+0.4$ & 0.250\\
\quad Mistral-Nemo-12B & $0.0$ & 6.0 & no & 1.000 & $-0.1$ & 0.625\\
\midrule
\multicolumn{7}{@{}l}{\emph{Sandwich, fixed in advance (fresh matched keys 12--19)}}\\
\quad Llama-3.1-8B & & & & & $\mathbf{-4.8}$ & \textbf{0.016}\\
\quad Qwen3.5-9B & & & & & $+0.3$ & 0.125\\
\bottomrule
\end{tabular}
\end{table}

\emph{Large effects.} Instructional prevention is attributed by the margin test on all three agents, as in Section~\ref{sec:defenses}, and it lowers the ASR under every matched key. Again much of the drop is inaction: averaged over the keys, the tool-action rate falls by 20 to 52 points.

\emph{Small effects.} The sandwich defense lowers the ASR of Llama-3.1-8B in every step: at two single seeds ($-4.5$ and $-1.8$ points), with disjoint keys ($-3.2$), and under each of the eight matched keys ($-4.6$ on average). The margin test does not attribute it, but the test fixed in advance confirms it: under each of the eight fresh keys the ASR is lower with the defense, by 4.8 points on average (adjusted $p=0.016$). For Qwen3.5-9B, the same test cannot distinguish the difference from zero ($+0.3$ points; adjusted $p=0.125$).

\emph{Why matched keys.} Even without a defense, the ASR under keys 0--5 exceeds that under keys 6--11, by 0.8 points for Qwen3.5-9B and by 1.6 for Llama-3.1-8B. Disjoint keys add such a chance difference to the effect; matched keys remove it (Appendix~\ref{app:averaged}) and suit small effects.

The count of attributed defenses in Section~\ref{sec:defenses} refers to the margin test, the stricter criterion, which refuses real effects such as this one. It does not say that the sandwich defense, or a defense that is not attributed on a given agent, has no effect; most of these combinations were not compared under keys.

\subsection{Stripped LLM gates}\label{sec:gatestrip}
We evaluate the stripped gate on the recorded decisions of the seven judges. A keyed order is a uniformly random order from the evaluator's side, so the recorded random-order presentations can serve as keyed presentations. For each proposal, we draw two disjoint sets of $K$ presentations and aggregate each set into a stripped decision. Over 500 draws, we record how often the two decisions are both decisive and different (a \emph{conflict}) and how often an evaluation abstains.

With one presentation per evaluation, the two evaluations conflict on 3--47\% of the proposals (Figure~\ref{fig:tare}b; Appendix~\ref{app:gate}). Requiring agreement among three presentations reduces conflicts to 0--2\% for every judge, which makes the gate consistent, though not necessarily right. The price is abstention, and its size differs between the judges: Gemini 3.8 Flash, Qwen3.8-27B, and K2-Horizon-375B abstain on 5--11\% of proposals, which a deployment can route to a human, and the four weaker judges on 38--65\%. At a given $K$ and threshold, the abstention rate thus reflects how strongly a judge depends on order. An online run with Llama-3.1-8B as the judge confirms Proposition~\ref{prop:gate} (Table~\ref{tab:practice}; Appendix~\ref{app:gate}).

Canonicalizing the evidence at the gate, for example by sorting its members, and asking the judge once also removes the requester's choice of order, at a third of the cost. Like a pinned seed, however, it fixes one order, and for 2--48\% of the proposals its decision differs from the one that most other orders produce; what the stripped gate adds is a signal of such order dependence, its abstentions. Neither defends against a requester who may also edit content and could resample presentations through trivial edits.

Table~\ref{tab:practice} compares \tare{} with current practice, benchmark by benchmark; the gains cost runs and abstentions.

\begin{table*}[t]
\caption{\tare{} compared with current practice, by benchmark: result and, in parentheses, method; the better result in bold. Planted effects: share of replicates in which the test finds the effect (0.0\% and 4.5\% without one). The last row gives the cost of the gains; the margin test has no row, because no current practice offers its guarantee.}\label{tab:practice}
\centering\footnotesize\setlength{\tabcolsep}{5pt}
\begin{tabular}{@{}L{6.35cm}L{4.45cm}L{4.98cm}@{}}
\toprule
What is measured & Current practice & \tare{}\\
\midrule
\multicolumn{3}{@{}l}{\emph{ASB, seven open-weight agents}}\\
Verdicts changed by the hash seed, of 400 & 7--59 (ASB's set order) & \textbf{0 for every agent} (keyed order)\\
ASR variance of null runs, vs.\ prediction & 1.7 times (global hash seed) & \textbf{0.9 times} (per-case keys)\\
Planted 3- and 2-point effects, Llama-3.1-8B & 29.5\%, 12.8\% (McNemar, one run per arm) & \textbf{100\%, 74.5\%} (randomization, 6 keys per arm)\\
Sandwich defense, Llama-3.1-8B (ASR points) & $-4.5$ or $-1.8$, by seed & \textbf{\boldmath$-4.8$, $p=0.016$} (8 fresh matched keys)\\
Runs with every request rejected, two agents & ASR 0\%: secure (two verdicts) & \textbf{not measured} (third verdict)\\
Delimiters on Mistral-Nemo-12B, 71 rejected & ASR 30.0\%, as undefended (two verdicts) & \textbf{\boldmath 30.3\% $\to$ 35.6\%} (third verdict, 320 cases)\\
\addlinespace[2pt]
\multicolumn{3}{@{}l}{\emph{ASB, Gemini 3.8 Flash and Gemini 4 Argon (production APIs)}}\\
Verdicts changed by an identical rerun & not measured (one run) & \textbf{13.8\% and 8.1\%} (same-state rerun)\\
Gemini 4 Argon runs ended by client crashes & 54 failed attacks (two verdicts) & \textbf{54 inconclusive} (third verdict)\\
\addlinespace[2pt]
\multicolumn{3}{@{}l}{\emph{AgentDojo, 35 seed-exposed cases}}\\
Llama-3.1-8B: outcomes changed by the seed & not varied (pinned seed) & \textbf{6} (seed varied; rerun changes 0)\\
K2-Horizon-375B: changed by an identical rerun & not measured (one run) & \textbf{10} (same-state rerun)\\
Gemini 3.8 Flash and Gemini 4 Argon: likewise & not measured (one run) & \textbf{0 and 0} (same-state rerun)\\
\addlinespace[2pt]
\multicolumn{3}{@{}l}{\emph{InjecAgent, Llama-3.1-8B, 397 cases}}\\
Verdicts changed by the serving instance & hidden (one instance) & \textbf{116} (rerun per instance)\\
\addlinespace[2pt]
\multicolumn{3}{@{}l}{\emph{LLM gates on 98 $\tau^2$-bench proposals}}\\
Contradictory decisions, seven judges & 3--47\% (judge asked once) & \textbf{0--2\%} (three keyed presentations)\\
Llama-3.1-8B judge: changed by submitted order & 41 of 98 (judge asked once) & \textbf{0 of 98} (stripped gate)\\
\midrule
Cost of a comparison; of a gate decision & 2 runs; 1 judge call & 12--16 runs; 3 judge calls, 5--65\% abstentions\\
\bottomrule
\end{tabular}
\end{table*}

\section{Discussion}\label{sec:discussion}

\subsection{What the results mean}\label{sec:meaning}
\emph{Reproducible is not the same as attributable.} The usual answer to unstable LLM evaluations, pinning the seed, serving the model deterministically, and rerunning, is necessary but not sufficient. Pinning makes one value of the content-free state reproducible, but it does not tell the reader how much another value would have changed, and on ASB that change is as large as many published defense effects. ASB's own paper reports 39 effects of its three prompt-level defenses on thirteen models, each as a single value without runs, seeds, or a variance~\cite{asb}; their median is 2.75 points in absolute value, and 32 of them are no larger than 5.0 points, the largest change of an agent's ASR that the hash seed alone produced in our runs. The models differ from ours, so the comparison only illustrates the problem.

\emph{The state is chosen, not only drawn.} Work on evaluation variance treats nuisance factors as random and reports a variance. For security claims this is not enough, because the parties who set the state also have an interest in the outcome. The clearest case is the LLM gate, where the requester who wants an action allowed also assembles its evidence, and for six of the seven judges, most proposals that the judge does not allow are allowed under some other order of the same evidence (Section~\ref{sec:judges}). A gate that has been evaluated under one order may therefore be weaker in deployment than its evaluation suggests. The stripped gate removes this choice and abstains where the judge depends on the order.

\subsection{Recommendations and cost}
The mechanism is not specific to ASB: any harness that derives the order of what the model sees from a hash, a set, or a sort on names exposes its verdicts to the same state, as AgentDojo does through the order of its simulated participants (Section~\ref{sec:seed}). \emph{Benchmark authors} should therefore not derive such orders from content-free state; a per-case evaluator key, recorded with the results, can order each case's items with a change of a few lines that leaves content and scoring untouched. Runs that fail for reasons unrelated to security need a third verdict instead of being counted as failed attacks. \emph{Evaluators} should report each result together with its declared content-free state, a same-state rerun on the serving instance used, the measured influence of each declared variable, and the inconclusive and tool-action rates; the \tare{} card (Appendix~\ref{app:artifact}) collects these on one page. When the expected effect is small, our ASB results suggest averaging over at least eight keys, using the same keys in both arms, and correcting for multiple comparisons. \emph{Builders of LLM gates} can present the evidence under several evaluator keys and abstain on disagreement; for the three most stable judges, a majority of three keyed presentations keeps conflicts at 2--5\% while abstaining on at most 1\% (Table~\ref{tab:gate}). The cost grows linearly with the number of keys: a run of ASB's 400 cases takes 0.5--2.4 GPU-hours on our hardware, a comparison with eight keys per arm 8--38, and a stripped gate with three presentations triples the cost of a decision.

\subsection{Limitations}\label{sec:limits}
\emph{Scope of the evidence.} Our deepest analysis concerns one benchmark, ASB, whose harness exposes the hash seed directly. AgentDojo provides a weaker dependence and InjecAgent a negative control, and the judges are tested on proposals from one source, which carry no reference decisions: we measure whether a gate is consistent, not whether it is right. Our defense analysis covers eighteen combinations on six agents, and Llama-3.1-70B has no defense runs; keyed-order runs have twelve keys for three agents and one key for the other four. A keyed order does not remove the dependence on the serving instance that some of our runs show, so our comparisons, and \tare{}'s guarantees, hold within one serving instance. The agents are open-weight models of up to 70B parameters, and the influences that we report are lower bounds from three to six seeds per agent (Section~\ref{sec:threat}). Through production APIs, whose serving floor exceeds the influence of the evaluator's state (Section~\ref{sec:seed}), our results do not say how much the seed matters, although the order effect on gates remains measurable there (Section~\ref{sec:judges}).

\emph{The tests.} The margin test is conservative and can refuse a real effect, such as that of the sandwich defense on Llama-3.1-8B (Section~\ref{sec:defense-keyed}). The randomization and matched tests have the coarse resolution of a few runs per arm, and of our averaged comparisons only the test under fresh keys was fixed entirely in advance (Section~\ref{sec:defense-keyed}). Proposition~\ref{prop:var} assumes independently varying cases, which per-case keys make plausible. Our own evaluator keys were drawn at random and are released with the runs, but were not committed in public before them.

\emph{Declared state only.} \tare{} measures and strips the state that an evaluator declares. Undeclared state, such as an ordering buried in a dependency, remains; a same-state rerun that is not identical signals it. Keyed orders apply only to collections that the claim treats as unordered (Section~\ref{sec:setting}), such as the members of an evidence object, and not to the turns of a conversation or to free text, whose order is content.

\section{Related Work}\label{sec:related}
\emph{Measurement validity.} Work on malware, fuzzing, ML-based security, and adversarial robustness shows how experimental design biases security conclusions and how to remove the bias~\cite{rossow2012prudent,tesseract,klees2018fuzz,arp2022dos,schloegel2024sok,carlini2019evaluating}. For agent security, Pathade et al.~\cite{pathade2026asr} find that most papers report neither variance nor repeated runs and that single runs misrank defenses, and Jia et al.~\cite{jia2026window} report that up to 35\% of single-run labels on AgentDojo change on a rerun. Others separate rerun from perturbation variation~\cite{noisefloor2026}, show that design choices within one harness shift a model's success by more than 30 points~\cite{componentbench2026}, and find scoring defects such as counting empty responses as successes~\cite{kapoor2024agents,zhu2025abc}. The statistical remedy, randomizing nuisance factors and reporting variance~\cite{henderson2018,bouthillier2021,miller2024errorbars}, is sound for state that a rerun redraws; we add choices made once in harness code, state that a party can select, and mechanisms that remove the state at its source. Studies of inference nondeterminism~\cite{atil2024nondet,song2024greedy} vary the server; we hold it deterministic and vary the harness. For ASB, Bhagwatkar et al.~\cite{bhagwatkar2025firewalls} observe that the benchmark forces the attacker tool into the plan; we measure the consequence of how the tool is inserted. Studies of over-defense~\cite{wang2026landscape,agentdyn2026} report the utility cost of defenses; we show, case by case, that an ASR reduction can consist mostly of inaction.

\emph{Order and presentation.} Position bias of LLMs is well documented in long contexts~\cite{liu2024lost}, multiple-choice options~\cite{pezeshkpour2024order,zheng2024mcq}, judges~\cite{zheng2023judge,wang2024fair,shi2024judging}, and tool selection~\cite{biasbusters2025,faghih2025tool}. Closer to security, TPRS~\cite{tprs2026} varies tool names and representations, Weng et al.~\cite{weng2026policy} rewrite the policies given to safety judges, and Zhou et al.~\cite{zhou2026style} wrap judged text in content-invariant styles against a repeat floor; all three add or rewrite text, whereas we only reorder and locate where harnesses set position from content-free state. Our stripped gate combines voting over samples~\cite{wang2023selfconsistency}, shuffling and aggregating~\cite{tang2024permutation}, and judging both orders~\cite{zheng2023judge}, and adds evaluator keys and a fail-closed abstention.

\emph{Invariance and noninterference.} Decision noninterference is an invariance requirement, like the metamorphic relations of software testing~\cite{chen2018metamorphic} and the invariance tests of CheckList~\cite{ribeiro2020checklist}; where those ask whether a model is robust, we ask whether a reported security effect survives the states that the evaluation pipeline can take. It adapts noninterference~\cite{goguen1982,biba1977,sabelfeld2003,clarkson2010hyper} to evaluation verdicts. Information-flow defenses, FIDES~\cite{fides2025} and CaMeL~\cite{camel2025}, control how untrusted content shapes actions, and \tare{} how content-free state shapes verdicts.

\section{Conclusion}\label{sec:conclusion}
Reproducibility and attribution are different properties. ASB's reruns are identical once the hash seed is fixed, yet the seed, a fixed tool order, the order of a judge's evidence, and the scoring of failed runs all change security verdicts. Decision noninterference states what a verdict should satisfy, and \tare{} makes it checkable and, where we tested it, removes the dependence: under evaluator-keyed orders the hash seed no longer changes any verdict. A security result on an LLM agent should therefore come with the measured influence of content-free state. And an attributed drop in ASR is not yet a security gain: much of every drop that we could attribute came from an agent that stopped acting, and one defense raised an agent's ASR beyond what content-free state explains.

\bibliographystyle{plainnat}
\bibliography{refs}

\appendix
\section{Artifact}\label{app:artifact}
The artifact (\url{https://anonymous.4open.science/r/TARE-1217}) contains the \tare{} package of Section~\ref{sec:tare}, with unit tests for the propositions; the scripts that produce this paper's ASB and judge numbers and figures from the raw run logs, with their outputs; the keyed-order patch for ASB and our other changes to its client as diffs against the pinned commit (Section~\ref{sec:setup}); the evaluator keys, the ledger of the keyed runs, and the design notes of the experiments whose analysis was fixed in advance; and the raw ASB result files and judge decisions (for AgentDojo and InjecAgent, the outputs of our comparison scripts). Our copies of ASB remove a two-second pause between requests and cache one environment query, and for the four agents whose chat templates reject a second system message the client merges ASB's system messages into one; none of these changes alters the content of a message. A script, \texttt{reproduce.sh}, runs the unit tests and every analysis script on the included raw results and compares each output with the expected one (all outputs agree on Linux and on macOS); \texttt{examples/} shows how to apply \tare{} to another benchmark and prints the \tare{} card of Llama-3.1-8B (Section~\ref{sec:procedure}).

\section{Guarantees}\label{app:guarantees}
Table~\ref{tab:guarantees} lists what each part of \tare{} guarantees and under which assumptions (Section~\ref{sec:tare}).

\emph{Proof of Proposition~\ref{prop:var}.} Two independent runs disagree on case $c$ with probability $2p_c(1-p_c)$; averaging over the cases gives $\bar I$. The frequency $f(c)$ is a mean of $K$ independent outcomes with success probability $p_c$, so its variance is $p_c(1-p_c)/K$. Independence across cases lets the variances add, which gives $\mathrm{Var}(\asr_K)=\frac{1}{|C|^2K}\sum_c p_c(1-p_c)$. \hfill$\square$

\emph{Per-case combinations.} If each case takes its verdict, for each system, from either of the two measured runs, the effect lies within $\bar\Delta\pm(m_A+m_B)/2$, where $\bar\Delta$ is the mean of the two measured effects, and it reaches both ends when every case whose verdict differs takes the verdict that moves the effect in one direction. Half the margin thus covers every per-case combination of the two measured states; the full margin is needed for an unmeasured state that changes as many verdicts but moves all of them against the effect.

\emph{Margin test for averages.} The test splits each system's $K$ runs into two halves, in the order in which the keys were drawn, and computes each case's success frequencies $f_X^{(1)}(c)$ and $f_X^{(2)}(c)$ in the halves, on the cases $C'$ conclusive in every run. The effect of a half is the difference between the systems' mean frequencies in it, and the margin is $\tilde I_A+\tilde I_B$ with $\tilde I_X=\frac{1}{|C'|}\sum_c|f_X^{(1)}(c)-f_X^{(2)}(c)|$. An effect is attributed if the effects of both halves exceed the margin in absolute value; the proof of Proposition~\ref{prop:bound} applies with frequencies in place of verdicts, so the guarantee covers alternative half-averages on the same cases. With $K=2$ this is the pair test, and with larger $K$ the margin tends to shrink.

\begin{table}[!htb]
\caption{Guarantees of \tare{} and their assumptions.}\label{tab:guarantees}
\centering\footnotesize\setlength{\tabcolsep}{5pt}
\begin{tabularx}{\textwidth}{@{}l>{\hsize=.8\hsize}Y>{\hsize=1.2\hsize}Y@{}}
\toprule
Part & Guarantee & Assumptions and scope\\
\midrule
Influence & lower bound on the largest influence of the state & none; covers only the values of the state that were tried\\
Margin test & an attributed effect keeps its sign at every state that moves at most as many verdicts as the measured change, in any direction & arms differ in nothing but the system and the varied state; no guarantee for states that move more verdicts\\
Randomization test & false detection at most $\alpha$ & null: the defense changes nothing; independent random keys (per run, or per pair if matched); runs executed alike on one deterministic instance or interchangeable ones\\
Variance (Prop.~\ref{prop:var}) & noise of an averaged ASR & independent keys and independently varying cases; checked empirically\\
Keyed order & DNI for the hash seed and the arrival order & the keyed collection is the only place where they reach the model\\
Stripped gate (Prop.~\ref{prop:gate}) & decision independent of the submitted order & deterministic judge; abstains\\
\bottomrule
\end{tabularx}
\end{table}

\section{All settings and full tables}\label{app:tables}
Tables~\ref{tab:seed}--\ref{tab:keyed} give the numbers behind Section~\ref{sec:findings}, Figures~\ref{fig:effects} and~\ref{fig:defense}, and Section~\ref{sec:kdraw}. Under batched serving, a same-seed rerun changes 1.4--6.3\% of the verdicts of four agents and a change of seed 6.7--15.1\%, so a stochastic server mixes its own variation into the seed's. Because inconclusive cases are excluded per comparison, case counts differ slightly between tables (for Mistral-Nemo-12B, 386--397). Two instances with identical flags can also differ: in the InjecAgent control, an instance that had compiled its computation graph and one that had loaded it from a cache differ in 116 of 397 verdicts at the same seed, and on a second machine the same seed changes 4 (Llama-3.1-70B) and 23 (Gemma-3-27B) of 400 verdicts. A likely mechanism, which we did not isolate, is that an instance selects its kernels when it compiles or loads its graph, so that two instances round the same logits differently; under greedy decoding, one changed token changes the rest of the transcript. Our comparisons stay within one instance or follow an exact same-state control. A comparison on conclusive cases is conditional on them: Mistral-Nemo-12B's delimiters run at seed 0 succeeds on 120 of its 329 conclusive cases, so its ASR over all 400 lies between 30.0\% and 47.8\%.

\emph{Statements of the abstract.} Its figures come from Table~\ref{tab:seed} (17\%: Llama-3.1-8B's largest pair; ASR range 5.0 points), Section~\ref{sec:seed} (no change at equal tool order), Table~\ref{tab:order} ($+18.8$, $+6.5$, $+6.6$), Table~\ref{tab:judges} (17 to 89 of 98 for six of the seven judges), Table~\ref{tab:defenses} and Figure~\ref{fig:defense} (attributed effects of instructional prevention and of delimiters), and Section~\ref{sec:defense-keyed} (the 4.8-point decrease). The 35--69\% are the shares of stopped attacks without a tool call in the attributed decreases: 69\% (76 of 110) for instructional prevention on Mistral-Nemo-12B at seed~0, 54\% on Qwen3.5-9B, 35\% on Llama-3.1-8B, and 58\% (32 of 55) for delimiters on Qwen3.5-9B.
\begin{table}[!htb]
\caption{The hash seed on ASB (400 cases); an identical rerun on the same instance changes no verdict. Two seeds: share of verdicts that differ between two seeds (mean, range over the pairs); all seeds: between any two of the agent's seeds.}\label{tab:seed}
\centering\footnotesize\setlength{\tabcolsep}{6pt}
\begin{threeparttable}
\begin{tabular}{@{}lcccc@{}}
\toprule
 & & \multicolumn{2}{c}{Verdicts changed (\%)} & ASR range\\
\cmidrule(lr){3-4}
Agent & Seeds & two seeds & all seeds & (points)\\
\midrule
Llama-3.1-8B & 5 & 14.2 (12.0--17.0) & 27.2 & 5.0\\
Llama-3.1-70B$^\ast$ & 3 & \phantom{0}4.0 (3.5--4.2) & \phantom{0}6.0 & 1.8\\
Qwen3-8B & 4 & \phantom{0}2.3 (1.8--2.8) & \phantom{0}4.2 & 1.5\\
Qwen3.5-9B & 6 & \phantom{0}2.5 (1.5--3.2) & \phantom{0}5.8 & 1.5\\
Qwen3-32B & 4 & \phantom{0}9.5 (8.0--10.5) & 16.8 & 1.8\\
Gemma-3-27B$^\ast$ & 4 & \phantom{0}6.0 (4.5--7.0) & 11.0 & 2.5\\
Mistral-Nemo-12B$^\dagger$ & 4 & \phantom{0}5.1 (3.8--6.3) & \phantom{0}8.9 & 2.5\\
\bottomrule
\end{tabular}
\begin{tablenotes}[flushleft]\scriptsize
\item[] $^\ast$Served on two GPUs; shown is the machine with more seeds. On the other machine, two seeds change 3.8\% (Llama-3.1-70B, one pair) and 5.3\% (Gemma-3-27B) of the verdicts. $^\dagger$The 395 cases that are conclusive at all four seeds.
\end{tablenotes}
\end{threeparttable}
\end{table}

\begin{table}[!htb]
\caption{Runs with failures, scored by ASB (two verdicts) and by \tare{} (third verdict; ASR on the run's conclusive cases). K2-Horizon-375B: four partial ASB runs through its API, every request an API error.}\label{tab:failed}
\centering\footnotesize\setlength{\tabcolsep}{5pt}
\begin{threeparttable}
\begin{tabular}{@{}lrrrr@{}}
\toprule
Run & cases & inconcl. & ASB & \tare{}\\
\midrule
K2-Horizon-375B API, 4 runs & 464 & 100\% & 0.0\% & not measured\\
Gemma-3-27B, unmodified client, 2 runs & 800 & 100\% & 0.0\% & not measured\\
Qwen3.5-9B, unmodified client, 2 runs & 800 & 100\% & 0.0\% & not measured\\
Mistral-Nemo-12B + delimiters, seed 0 & 400 & 17.8\% & 30.0\% & 36.5\%\\
Gemini 4 Argon API, seed 0$^\ast$ & 400 & 4.0\% & 18.0\% & 18.8\%\\
\bottomrule
\end{tabular}
\begin{tablenotes}[flushleft]\scriptsize
\item[] $^\ast$16 content-filter crashes in this run (54 in three); silent truncations leave no trace.
\end{tablenotes}
\end{threeparttable}
\end{table}

\begin{table}[!htb]
\caption{Sorted versus reverse-sorted tool list, seed 0 (ASR in \%; set: ASB's set order). Only s./r.: cases that succeed only when sorted / reverse-sorted; Holm $p$: exact McNemar test; scen.: scenarios with a higher / lower ASR when sorted (sign-flip test, Holm-adjusted $p=0.027$ and $0.047$ for the Llama agents, $0.41$ for Mistral-Nemo-12B).}\label{tab:order}
\centering\footnotesize\setlength{\tabcolsep}{5pt}
\begin{tabular}{@{}lrrrrrrr@{}}
\toprule
Agent & sorted & reverse & diff. & set & only s./r. & Holm $p$ & scen.\\
\midrule
Llama-3.1-8B & 48.2 & 29.5 & $+18.8$ & 36.2 & 84 / 9 & $2\cdot10^{-15}$ & 9 / 1\\
Llama-3.1-70B & 25.2 & 18.8 & $+6.5$ & 22.0 & 28 / 2 & $5\cdot10^{-6}$ & 8 / 0\\
Mistral-Nemo-12B & 35.7 & 29.1 & $+6.6$ & 30.6 & 33 / 7 & $2\cdot10^{-4}$ & 6 / 2\\
Qwen3-8B & 9.0 & 11.0 & $-2.0$ & 9.8 & 3 / 11 & 0.2 & 0 / 5\\
Qwen3.5-9B & 20.5 & 21.2 & $-0.8$ & 19.8 & 8 / 11 & 0.8 & 4 / 5\\
Qwen3-32B & 62.8 & 60.2 & $+2.5$ & 58.2 & 36 / 26 & 0.8 & 6 / 2\\
Gemma-3-27B & 10.0 & 8.2 & $+1.8$ & 8.2 & 24 / 17 & 0.8 & 7 / 1\\
\bottomrule
\end{tabular}
\end{table}

\begin{table}[!htb]
\caption{Seven LLM judges on 98 proposals (Gemini 3.8 Flash and K2-Horizon-375B through APIs). Split: the decision differs across orders (canonical and ten random ones; K2-Horizon-375B eight, Gemini 3.8 Flash six), of which only ALLOW/ASK or only ALLOW/BLOCK. Floor: differs between identical repeats (at most 2.8\% of the pairs of repeats). Canon.\ $\ne$ maj.: canonical decision differs from the random-order majority (ties excluded). $\to$ALLOW: allowed in a random but not the canonical order. $^\ast$Online: 0 of 294 identical pairs differ.}\label{tab:judges}
\centering\footnotesize\setlength{\tabcolsep}{5pt}
\begin{tabular}{@{}lrrrrrr@{}}
\toprule
Judge & split & \shortstack{ALLOW\\/ASK} & \shortstack{ALLOW\\/BLOCK} & floor & \shortstack{canon.\\$\ne$ maj.} & \shortstack{$\to$\\ALLOW}\\
\midrule
Gemini 3.8 Flash & 7 (7\%) & 0 & 7 & 1/98 & 2/98 & 2/6\\
Qwen3.8-27B & 17 (17\%) & 0 & 17 & 0/98 & 3/97 & 4/5\\
K2-Horizon-375B & 25 (26\%) & 0 & 25 & 4/97 & 6/96 & 6/8\\
Gemma-3-27B & 64 (65\%) & 59 & 5 & 0/98 & 18/92 & 34/41\\
Qwen3.5-9B & 87 (89\%) & 3 & 70 & 2/98 & 14/86 & 55/77\\
Llama-3.1-8B & 88 (90\%) & 72 & 11 & 0/98$^\ast$ & 23/90 & 44/51\\
Mistral-Nemo-12B & 89 (91\%) & 76 & 10 & 0/98 & 42/88 & 47/58\\
\bottomrule
\end{tabular}
\end{table}

\begin{table}[!htb]
\caption{The attribution test for ASB's defenses on the cases conclusive in all four runs (400; Mistral-Nemo-12B 390, 395, 320; Gemma-3-27B with delimiters 344), seeds 0 and 1 (Gemma-3-27B: 0 and 2); effects and margin in points. $^\dagger$Follow-up combination, run under a design fixed in advance on a new serving instance, with an undefended control at the same seed as base arm (Appendix~\ref{app:n15}). $^\circ$Not attributed, but the mean effect exceeds half the margin (inner band of Figure~\ref{fig:defense}), so the sign survives every per-case combination of the seeds. No call: stopped attacks without a tool call; task: benign task success (\%), undefended$\to$defended, first seed; instr.: instructional prevention; delim.: delimiters.}\label{tab:defenses}
\centering\footnotesize\setlength{\tabcolsep}{5pt}
\begin{tabular}{@{}llrrccr@{}}
\toprule
Agent & Defense & effect & margin & attr. & no call & task\\
\midrule
Mistral-Nemo-12B & instr. & $-$26.2/$-$25.1 & 6.2 & yes & 69\% & 2.3$\to$2.3\\
 & sandwich & $-$0.3/$-$0.3 & 9.6 & no & 0\% & 2.3$\to$6.8\\
 & delim. & $+$5.3/$+$5.6 & 9.7 & no$^\circ$ & 0\% & 0.9$\to$16.2\\
\addlinespace[1pt]
Qwen3.5-9B & instr. & $-$10.0/$-$8.5 & 5.0 & yes & 54\% & 33.5$\to$18.0\\
 & sandwich & $+$0.8/$-$0.5 & 6.8 & no & 0\% & 33.5$\to$36.0\\
 & delim.$^\dagger$ & $-$8.2/$-$8.5 & 5.2 & yes & 58\% & 33.5$\to$14.8\\
\addlinespace[1pt]
Gemma-3-27B & instr. & $+$2.2/$+$3.2 & 11.5 & no & 43\% & 11.0$\to$15.2\\
 & sandwich & $+$0.2/$+$0.2 & 9.0 & no & -- & 11.0$\to$12.2\\
 & delim.$^\dagger$ & $+$1.5/$+$2.6 & 11.0 & no & 29\% & 10.5$\to$7.6\\
\addlinespace[1pt]
Qwen3-32B & instr. & $+$1.0/$+$2.5 & 19.5 & no & 0\% & 27.5$\to$19.5\\
 & sandwich & $-$18.5/$-$16.0 & 21.0 & no$^\circ$ & 0\% & 27.5$\to$29.0\\
 & delim.$^\dagger$ & $-$15.2/$-$14.0 & 18.8 & no$^\circ$ & 0\% & 26.2$\to$21.8\\
\addlinespace[1pt]
Llama-3.1-8B & instr. & $-$23.0/$-$21.2 & 20.2 & yes & 35\% & 50.5$\to$17.0\\
 & sandwich & $-$4.5/$-$1.8 & 25.8 & no & 0\% & 50.5$\to$54.8\\
 & delim. & $-$14.8/$-$11.8 & 24.5 & no$^\circ$ & 31\% & 50.5$\to$19.0\\
\addlinespace[1pt]
Qwen3-8B & instr.$^\dagger$ & $-$4.2/$-$3.2 & 5.0 & no$^\circ$ & 90\% & 8.0$\to$1.8\\
 & sandwich$^\dagger$ & $-$3.0/$-$2.2 & 3.2 & no$^\circ$ & 0\% & 8.0$\to$7.5\\
 & delim.$^\dagger$ & $+$18.0/$+$20.5 & 8.0 & yes & 14\% & 8.0$\to$13.5\\
\bottomrule
\end{tabular}
\end{table}

\begin{table}[!htb]
\caption{Keyed tool order on ASB, three agents with twelve keys. Top: changed verdicts between two runs (of 400; Mistral-Nemo-12B 395--397). Middle: influence; SD of the ASR, observed / predicted by Proposition~\ref{prop:var}. Bottom: margin of the averaged test, same agent in both arms (mean of 60 key sets).}\label{tab:keyed}
\centering\footnotesize\setlength{\tabcolsep}{6pt}
\begin{tabular}{@{}lrrr@{}}
\toprule
 & Llama-3.1-8B & Nemo-12B & Qwen3.5-9B\\
\midrule
ASB set order, seed 0 vs.\ 1 & 59 & 19 & 11\\
same key, seed 0 vs.\ 1 & 0 & 0 & 0\\
same key, other instance or node & 0, 0 & -- & 0\\
same key, other node, 4 days later & 0 & 0 & 0\\
\midrule
influence between keys & 13.6\% & 5.1\% & 2.8\%\\
\quad between hash seeds & 14.2\% & 5.1\% & 2.5\%\\
SD of ASR across keys & 1.21 / 1.31 & 0.66 / 0.81 & 0.67 / 0.59\\
\quad across hash seeds & 2.09 / 1.33 & 1.19 / 0.80 & 0.66 / 0.56\\
\midrule
null margin, $K=2$ & 27.1 & 10.3 & 5.5\\
\quad $K=4$ & 21.6 & 7.8 & 4.4\\
\quad $K=6$ & 17.6 & 6.8 & 3.9\\
\bottomrule
\end{tabular}
\end{table}

\section{Follow-up defense combinations}\label{app:n15}
Of the eighteen combinations of agent and defense in Section~\ref{sec:defenses}, twelve were measured first. The other six ($\dagger$ in Table~\ref{tab:defenses} and Figure~\ref{fig:defense}) were run afterwards: delimiters on Gemma-3-27B and Qwen3.5-9B, whose first runs had every request rejected (Section~\ref{sec:failed}), delimiters on Qwen3-32B, and all three defenses on Qwen3-8B. A design note written before any of their runs (N15, in the artifact) fixed the runs, their analysis, and that every result is reported, with the statistics of the first twelve. Each comparison stays within one new serving instance, with an undefended control on that instance as its base arm; two delimiters runs of Qwen3.5-9B that our scheduler cut short were repeated with new controls, as the note records. We did not run Llama-3.1-70B, whose runs need about two thousand model calls each on two GPUs. Its seeds change 4.0\% of its undefended verdicts (Table~\ref{tab:seed}); in the eighteen comparisons, the margin was 1.3 to 4.6 times the undefended agent's influence, which would put its margins at about 5 to 18 points.

\emph{Checks.} The controls repeat earlier undefended runs at the same seeds: for Qwen3-8B and Qwen3.5-9B they reproduce them verdict for verdict, also on the workstation GPU; for Gemma-3-27B, 19 and 16 of the 400 verdicts differ, and for Qwen3-32B, 52 and 63. Under delimiters, Gemma-3-27B's chat template rejects the same 56 cases at both seeds, whose messages do not alternate between user and assistant; \tare{} marks them inconclusive, where ASB would count failed attacks (Section~\ref{sec:failed}).

\emph{Results.} Section~\ref{sec:defenses} discusses these six combinations together with the first twelve. Among them, \tare{} attributes two effects, both of delimiters and in opposite directions. Both survive the worst treatment of the excluded cases, since all 400 cases are conclusive in both comparisons, and the margin from the largest seed pair (5.8 and 9.0 points). The sandwich defense on Qwen3-8B and delimiters on Gemma-3-27B and Qwen3-32B stay within their margins.

\section{Null pairs and planted effects}\label{app:planted}
\emph{Null pairs.} Two runs of the same undefended agent at different seeds are a comparison of two identical systems, so a significant difference is a false positive. An exact McNemar test at the 5\% level~\cite{dietterich1998} rejects for 3 of the 52 such pairs, which neither shows nor excludes miscalibration, because the pairs of an agent share runs; under per-case keys, 3 of the 198 pairs of keyed runs of the three agents with twelve keys differ significantly.

\emph{Planted effects.} Table~\ref{tab:plantedk} gives the planted-effect experiment of Section~\ref{sec:planted} for all settings. The rates without a planted effect do not estimate a test's level: all replicates of an agent are splits of one pool of twelve runs. The artifact also has the experiment for single runs under ASB's set order. There, without a planted effect, McNemar's test reports an effect in 22.3\% of the replicates for Llama-3.1-8B and in 17.3\% for Mistral-Nemo-12B, because some of their seed pairs differ significantly. No procedure reported an effect of the wrong sign.

\begin{table}[!htb]
\centering\footnotesize\setlength{\tabcolsep}{5pt}
\caption{Planted defense effects, averaged keyed runs with $K$ keys per arm (2,000 replicates per cell): replicates (\%) in which a procedure reports an effect, by planted effect (points; at zero every report is false). Randomization and McNemar tests at the 5\% level.}\label{tab:plantedk}
\begin{tabular}{@{}llrrrrrr@{}}
\toprule
Agent ($K$) & Test & 0 & 1 & 2 & 3 & 5 & 10\\
\midrule
Llama-3.1-8B (4) & randomization & 2.1 & 10.0 & 27.6 & 65.0 & 100 & 100\\
 & McNemar, one run & 0.0 & 4.6 & 14.8 & 30.6 & 68.9 & 100\\
 & margin test & 0.0 & 0.0 & 0.0 & 0.0 & 0.0 & 0.0\\
\addlinespace[2pt]
Llama-3.1-8B (6) & randomization & 4.5 & 21.2 & 74.5 & 100 & 100 & 100\\
 & McNemar, one run & 0.0 & 4.8 & 12.8 & 29.5 & 69.2 & 100\\
 & margin test & 0.0 & 0.0 & 0.0 & 0.0 & 0.0 & 0.0\\
\addlinespace[2pt]
Mistral-Nemo-12B (6) & randomization & 3.0 & 63.0 & 100 & 100 & 100 & 100\\
 & McNemar, one run & 0.0 & 4.4 & 24.4 & 53.5 & 98.3 & 100\\
 & margin test & 0.0 & 0.0 & 0.0 & 0.0 & 3.6 & 100\\
\addlinespace[2pt]
Qwen3.5-9B (6) & randomization & 3.4 & 63.2 & 100 & 100 & 100 & 100\\
 & McNemar, one run & 5.4 & 15.4 & 39.4 & 74.9 & 99.8 & 100\\
 & margin test & 0.0 & 0.0 & 0.2 & 14.0 & 99.2 & 100\\
\bottomrule
\end{tabular}
\end{table}

\section{Stages of the averaged comparisons}\label{app:averaged}
\emph{First stage.} The comparisons of Section~\ref{sec:defense-keyed} first ran with four keys per arm (defended 0--3, undefended 4--7). Instructional prevention changed the ASR by $-25.3$, $-9.8$, and $-27.6$ points on Llama-3.1-8B, Qwen3.5-9B, and Mistral-Nemo-12B, all attributed by the margin test, and the sandwich defense by $-3.7$, $+0.8$, and $-0.1$, none attributed. With four keys the randomization test cannot fall below $p=0.029$, which no comparison passes after Holm's correction over six, so we fixed the six-key design and extended all comparisons irrespective of their results (for the sandwich defense on Llama-3.1-8B all twelve runs existed already). Every split of an arm's six keys into halves gives the same attributions.

\emph{Matched keys.} Table~\ref{tab:matched} gives the per-key differences. With six keys the matched test cannot fall below $p=0.031$, whereas with eight its smallest value, 0.008, can pass the correction over six comparisons. We therefore fixed the eight-key design in a note and ran the defended arms under keys 6 and 7, after each agent's new serving instance had reproduced the undefended run under key 0 exactly.

\emph{Decomposition.} On the same cases, an effect under disjoint keys is the sum of the matched effect under keys 0--5 and the difference between the undefended runs under keys 0--5 and 6--11: $-3.21=-4.83+1.62$ for Llama-3.1-8B and $+1.25=+0.50+0.75$ for Qwen3.5-9B. Such differences between two sets of six keys are unusual (unadjusted $p=0.013$ and $0.06$; Mistral-Nemo-12B: 0.1 points, $p=0.92$); all agents share the keys, so these are not independent.

\emph{Test fixed in advance.} A further note, written before any of its runs, specifies the third step. We chose its two comparisons, and the direction expected for Llama-3.1-8B, with the earlier results in hand; the other four comparisons are not tested again. The note fixes both arms under keys 12--19, under which no run of any agent existed, the two runs of a key on one serving instance, the matched test over the eight differences, and Holm's correction over the two comparisons. The effect on Llama-3.1-8B counts as confirmed if the mean difference is negative and the adjusted two-sided p-value is below 0.05; for Qwen3.5-9B the note makes no prediction. Each serving instance (two for Llama-3.1-8B, three for Qwen3.5-9B) first repeated the undefended run under key 0, identically in all 400 outcomes. All 32 runs completed, none was repeated, and both runs of a key share one instance. As the note prescribes, we also give the untested mean over all sixteen keys: $-4.7$ points for Llama-3.1-8B (lower under each) and $+0.3$ for Qwen3.5-9B.

\begin{table}[!htb]
\caption{Matched keys: defended minus undefended ASR under the same key, in points; right: test fixed in advance (sandwich, keys 12--19). $-$/$+$: keys with a lower / higher ASR; $p$: matched test; Holm: over six and over two comparisons.}\label{tab:matched}
\centering\footnotesize\setlength{\tabcolsep}{4pt}
\begin{tabular}{@{}lrrrrrr@{\hspace{8pt}}lrr@{}}
\toprule
 & \multicolumn{2}{c}{Llama-3.1-8B} & \multicolumn{2}{c}{Qwen3.5-9B} & \multicolumn{2}{c}{Nemo-12B} & & \multicolumn{2}{c}{fixed in advance}\\
\cmidrule(lr){2-3}\cmidrule(lr){4-5}\cmidrule(lr){6-7}\cmidrule(l){9-10}
Key & instr. & sand. & instr. & sand. & instr. & sand. & Key & Llama & Qwen\\
\midrule
0 & $-26.0$ & $-3.8$ & $-9.0$ & $+1.0$ & $-27.7$ & $0.0$ & 12 & $-4.0$ & $+1.0$\\
1 & $-27.0$ & $-6.0$ & $-10.0$ & $0.0$ & $-29.0$ & $-0.3$ & 13 & $-4.5$ & $+0.5$\\
2 & $-25.5$ & $-4.8$ & $-11.5$ & $+1.0$ & $-27.2$ & $0.0$ & 14 & $-5.5$ & $-0.2$\\
3 & $-28.0$ & $-5.5$ & $-9.5$ & $+0.5$ & $-26.7$ & $0.0$ & 15 & $-4.8$ & $+0.5$\\
4 & $-25.5$ & $-5.2$ & $-10.2$ & $-0.2$ & $-26.2$ & $+0.3$ & 16 & $-4.8$ & $0.0$\\
5 & $-27.5$ & $-3.8$ & $-9.0$ & $+0.8$ & $-26.2$ & $-0.3$ & 17 & $-5.5$ & $0.0$\\
6 & $-26.5$ & $-4.2$ & $-8.5$ & $-0.2$ & $-27.5$ & $-0.3$ & 18 & $-6.0$ & $0.0$\\
7 & $-24.8$ & $-3.8$ & $-10.0$ & $+0.2$ & $-28.0$ & $0.0$ & 19 & $-3.5$ & $+0.8$\\
\midrule
mean & $-26.3$ & $-4.6$ & $-9.7$ & $+0.4$ & $-27.3$ & $-0.1$ & & $-4.8$ & $+0.3$\\
$-$/$+$ & 8/0 & 8/0 & 8/0 & 2/5 & 8/0 & 3/1 & & 8/0 & 1/4\\
$p$ & 0.008 & 0.008 & 0.008 & 0.125 & 0.008 & 0.625 & & 0.008 & 0.125\\
Holm & 0.047 & 0.047 & 0.047 & 0.250 & 0.047 & 0.625 & & 0.016 & 0.125\\
\bottomrule
\end{tabular}
\end{table}

\section{Stripped gates: all settings}\label{app:gate}
Table~\ref{tab:gate} gives the stripped gate of Section~\ref{sec:gatestrip} for other $K$ and $\tau$. In the online run with Llama-3.1-8B as the judge ($K=3$, $\tau=1$), each proposal was submitted in two random orders under one set of three keys and once more under a second set. The stripped decision was the same for both submitted orders of every proposal, whereas the judge asked directly gave different decisions for 41; the two key sets conflicted on 3 proposals, and the gate abstained in 49.5\% of the 196 stripped decisions (offline estimates: 1\% and 57\%).
\begin{table}[!htb]
\caption{Stripped gate: conflicts between two independent evaluations of a proposal / abstention rate. $K=1$: unstripped judge; $\tau$: share of the $K$ answers that must agree; n/a: too few recorded orders.}\label{tab:gate}
\centering\footnotesize\setlength{\tabcolsep}{6pt}
\begin{tabular}{@{}lrrrr@{}}
\toprule
 & $K=1$ & $K=3$, $\tau=0.51$ & $K=3$, $\tau=1$ & $K=5$, $\tau=1$\\
Judge & conflict & confl./abst. & confl./abst. & confl./abst.\\
\midrule
Gemini 3.8 Flash & 3\% & 2\% / 0\% & 0\% / 5\% & n/a\\
Qwen3.8-27B & 7\% & 5\% / 0\% & 0\% / 10\% & 0\% / 13\%\\
K2-Horizon-375B & 8\% & 4\% / 1\% & 0\% / 11\% & n/a\\
Gemma-3-27B & 26\% & 17\% / 1\% & 1\% / 38\% & 0\% / 51\%\\
Qwen3.5-9B & 47\% & 28\% / 10\% & 1\% / 65\% & 0\% / 80\%\\
Llama-3.1-8B & 39\% & 28\% / 1\% & 1\% / 57\% & 0\% / 75\%\\
Mistral-Nemo-12B & 39\% & 28\% / 1\% & 2\% / 57\% & 0\% / 80\%\\
\bottomrule
\end{tabular}
\end{table}

\end{document}